\documentclass[preprint,showpacs,preprintnumbers,amssymb,aps]{revtex4}

\usepackage{graphicx}
\usepackage{dcolumn}
\usepackage{bm}

\begin{document}


\title{Relativistic Mean Field Model with 
Generalized Derivative Nucleon-Meson Couplings}

\author{S. Typel}
\affiliation{
Department of Physics and Astronomy and
National Superconducting Cyclotron Laboratory,
Michigan State University, 
East Lansing, Michigan 48824-1321
}

\author{T. v. Chossy}
\author{H. H. Wolter}
\affiliation{
Sektion Physik, Universit\"{a}t M\"{u}nchen, Am Coulombwall 1,
D-85748 Garching, Germany}

\date{\today}

\begin{abstract}
The quantum hadrodynamics (QHD) model with minimal nucleon-meson
couplings is generalized by introducing couplings of mesons to 
derivatives of the nucleon field in the Lagrangian density. 
This approach allows
an effective description of a state-dependent in-medium interaction
in the mean-field approximation.
Various parametrizations for the generalized couplings
are developed and applied to
infinite nuclear matter. In this approach, scalar and vector 
nucleon self-energies
depend on both density and momentum similarly as in the Dirac-Brueckner
theory. The Schr\"{o}dinger-equivalent optical potential 
is much less repulsive at high nucleon energies as compared to standard
relativistic mean field models and thus agrees better with 
experimental findings. The derivative couplings in the extended model
have significant effects on the properties of symmetric nuclear 
matter and neutron matter.
\end{abstract}

\maketitle

\section{Introduction}

Properties of nuclear matter and finite nuclei have been described with great
success by relativistic mean field (RMF) models in recent years 
\cite{Wal74,Rin96,Ser86,Ser92,Gam90,Ruf88,Rei86,Hor81,Chi77}. These models
provide a novel saturation mechanism, an explanation of the strong spin-orbit
interaction, and a natural energy dependence of the optical potential. 
They are often referred to as quantum hadrodynamics (QHD) since nuclear systems
are described in these quantum field theories in terms of interacting hadrons,
i.e.\ nucleon and meson fields. 
Lagrangians have been proposed in simple versions
about 25 years ago \cite{Wal74}, 
and since then there have been many different treatments,
extensions and applications.

There are basically two approaches of QHD, which one may call microscopic
and phenomenological, respectively. The microscopic method is the
Dirac-Brueckner (DB) approach 
\cite{deJ98,Seh97,Hub95,Boe94a,Boe94b,Fri94,Li92,deJ91,Bro90,Bot90,%
Mue88,Hor87,tHM87,Ana83}
which tries to connect nucleon-nucleon scattering
with finite nuclear systems via many-body theory. The effective Lagrangian
has a simple form where nucleons and mesons couple in a minimal way.
The coupling constants are fitted to NN scattering observables
via the relativistic T-matrix approach. Then nuclear matter is treated
in the Brueckner approximation and an effective in-medium interaction
is extracted. The relativistic Brueckner approach successfully achieves
saturation for nuclear matter which was not possible in non-relativistic
methods with only two-body forces. Nucleon self energies $\Sigma_{i}$
($i=$ scalar, vector, tensor, \dots) in the 
DB approach are both density and momentum dependent. Below
the Fermi momentum $k_{F}$ the momentum dependence is found to be not
very strong  but above $k_{F}$ it corresponds to the momentum
dependence in the empirical optical potential. While the DB method
is very successful in describing infinite nuclear matter, it is very
difficult to apply to finite nuclei without drastic approximations,
e.g. the local density approximation (LDA).
In order to describe detailed properties of finite nuclei
some adjustements have to be introduced \cite{Boe94b,Hof01,Had93,Bro92}.

In the phenomenological approach QHD is applied directly to nuclear systems
without reference to NN scattering. Simple many-body approximations,
usually the mean field approximation without exchange and correlation
contributions, are employed. The coupling strengths of the mesons
are directly adjusted to saturation properties of nuclear matter.
In the original approach proposed by Walecka \cite{Wal74}
the self energies are proportional to the corresponding densities.
It was found that in this approach with linear meson-nucleon couplings
properties of nuclear matter are not satisfactory away from saturation
(too large incompressibility and too small effective mass) and
finite nuclei are not described well.
In order to improve the description the QHD Lagrangian has to be extended
to include medium and state dependent effects in an effective way.
This has been achieved by adding non-linear and higher order interaction
terms to the Lagrangian or assuming density-dependent couplings 
of the nucleon and meson fields.

Non-linear models are the most prominent approach, in which cubic and
quartic self-interactions of the $\sigma$-meson are introduced.
This extension was proposed by Boguta and Bodmer
\cite{Bog77} in order to lower the incompressibility of nuclear matter.
Later the approach was extended to other meson fields \cite{Bod91}.
Various parametrizations have been introduced which provide a good
description of a whole range of nuclei
\cite{Lal97,Sug94,Ruf88,Rei86,Sha00}. Replacing the minimal coupling of the
$\sigma$-meson to the nucleon by a derivative coupling, Zimanyi and
Moszkowski (ZM) obtained a particular non-linear $\sigma$-coupling
after rescaling of the nucleon field \cite{Zim90}. Although this approach
leads to a small incompressibility of nuclear matter, the effective nucleon 
mass turns out to be rather large, and, correspondingly, 
the spin-orbit interaction
is too small. The ZM model with further extensions
was studied in detail in \cite{Mit02,vCh00,Del94}.
Recently, this class of approaches has
been put on a more systematic basis in the effective field theories (EFT)
where the possible terms and their importance are systematically
categorized \cite{Fur00b,Fur98,Fur97,Del01,Del99}.

An alternative approach assumes a density dependence of the
coupling vertices \cite{Len95,Fuc95,Typ99}. 
This extension is suggested by DB theory. 
However, in a thermodynamically consistent and Lorentz covariant
model, the couplings cannot 
depend parametrically on the density. They have to be lorentz-scalar 
functionals of the field operators. Then the Euler-Lagrange variational
equations lead to so-called rearragement contributions in the self-energies. 
The first serious parametrization for finite nuclei 
was developed in \cite{Typ99}
where a dependence of the couplings on the vector density (VDD) was found to 
work best.
It was shown that good results can be obtained in the whole chart
of nuclei, comparable to the best non-linear models.
Recently, a new parametrization within the VDD approach
has been developed \cite{Nik02}. 
These models can be considered in the general framework of
density-functional theory which is given here in terms of the
density dependent vertex functions. This connection is
rather close, since the density dependence is 
motivated by DB theory which includes exchange and correlation terms.

Generally it seems desirable to describe the structure of nuclei and
nuclear reactions in a common framework. This implies that one also
needs the momentum dependence of the self energies at least for momenta above
the Fermi surface. Empirically these are well studied in 
Dirac phenomenology  by fitting scalar and vector potentials 
to elastic proton nucleus 
scattering data \cite{Ham90,Coo93}. It is found that the
optical potential becomes repulsive with increasing energy
but levels off at about 1~GeV.
Optical potentials extracted from standard RMF models
are much too large and increase linearly at high energies. 
Recently, a reasonable description
of proton-nucleus scattering and an improved optical potential
has been obtained by multiplying the self-energies from RMF calculations
of finite nuclei with functions that depend explicitly
on the proton energy \cite{Typ02}. However, the consistency of the
theory is lost and the change of the energy dependence at different
medium densities is not properly taken into account in this approach.
The momentum dependence of the interaction also becomes important
in the description of heavy-ion collisions. Several
parametrizations for a parametric momentum dependence of the potentials
have been developed in different approaches
\cite{Dan00,Lee97,Li93,Mar94,Web93,Web92}.

So far, extensions of the RMF models concentrated on the
effective density dependence of the in-medium interaction
since they were mostly applied to the description of finite nuclei.
If one goes beyond
the mean field approximation momentum dependent self energies will
emerge naturally. Recently, it was
shown, that an energy-dependence of the self-energies can
improve the level density close to the Fermi surface of finite
nuclei \cite{Vre02}. In this RMF model, dynamical effects 
from the coupling of the single particle motion 
to collective surface vibrations
are taken into account in a phenomenological approach. However,
the energy-dependence was introduced only parametrically into 
the self-energies destroying the covariance of the theory.
In order to 
retain the simple structure of the equations known from the
RMF approximation one can imagine that it is desirable to modify
the QHD Lagrangian in such a way that it leads to 
momentum dependent self energies already in the mean field approximation. 
In this work we propose such a class of Lagrangians 
with generalized derivative meson-nucleon
couplings. The various types of possible couplings will
have characteristic consequences in the field equations.
Here we limit ourselves to the study of infinite nuclear matter within
the derivative coupling (DC) 
model in order to demonstrate the main effects and the limitations 
of the approach. The results can serve as a foundation for a 
later application of the model to finite nuclei. 
Although derivative couplings have been investigated in simple approaches
before, they were essentially cast into density-dependent forms, 
e.g.\ by rescaling
the fields as in the ZM model.
In EFT models contributions containing derivatives of the density
are included, but generally they do not lead to a momentum
dependence of the self energies and they do not affect properties of nuclear
matter.

The paper is organized as follows. In Sec.\ \ref{Sec2} the
the Lagrangian of the DC model
is introduced and the field equation for mesons
and nucleons are derived in the mean field approximation.
In the following section the DC model is applied to infinite nuclear matter
and expressions for basic quantities are derived.
Three parametrizations are developed in Sec.\ \ref{Sec4} and
the resulting properties of symmetric and asymmetric nuclear matter
are compared to standard RMF models.
Finally, we summarize and give an outlook in Sec.\ref{Sec5}.

\section{Lagrangian density and field equations}
\label{Sec2}

The description of nuclear systems in QHD starts from
a Lagrangian density $\mathcal{L}$
that contains nucleons and mesons as degrees of freedom.
In case of symmetric nuclear matter it is sufficient
to include isoscalar $\sigma$- and $\omega$-mesons. However, for 
asymmetric matter and finite nuclei, isovector mesons have to be added. 
It is customary to consider only the vector $\rho$-meson but for the sake of
completeness we also take the scalar $\delta$-meson into account.
The standard approach is to couple the nucleons minimally to
the mesons. In case of the isoscalar
mesons this leads to vertices of the form $\bar{\psi} \psi \sigma$
and $\bar{\psi} \gamma^{\mu}\psi \omega_{\mu}$  
corresponding to the linear Walecka model.

The idea of the DC model is to introduce 
explicit couplings of the mesons to the first derivative of the
nucleon field $\psi$, e.g. $\bar{\psi} i \partial_{\mu} \psi$ or
$\bar{\psi} \gamma^{\mu} i \partial_{\mu} \psi$. 
The derivate $i\partial_{\mu}$ in the new coupling vertices is
actually replaced  by the covariant derivative $iD_{\mu}$
as it appears in the usual kinetic contribution to the Lagrangian.
In the current model we consider couplings to the covariant derivative of the
nucleon field which are linear or quadratic in the
$\sigma$- and $\omega$-meson fields.
For simplicity for the $\delta$- and $\rho$-mesons
only linear couplings are taken into account.
Of course, the standard minimal couplings are retained in the DC model.

On these conditions the Lagrangian density in the DC model assumes the form
\begin{equation} \label{lagdef}
 \mathcal{L} = 
 \frac{1}{2} \left[  \bar{\psi}  \Gamma_{\mu} i D^{\mu} \psi
 + \overline{(i D^{\mu}\psi)}  \Gamma_{\mu}  \psi \right]
 -  \bar{\psi} \left(m-\Gamma_{\sigma}\sigma
  -\Gamma_{\delta} \vec{\tau} \cdot \vec{\delta} \right) \psi 
 + {\cal L}_{m}
\end{equation}
with the covariant derivative
\begin{equation}
 iD_{\mu} = 
 i \partial_{\mu} -\Gamma_{\omega} \omega_{\mu}
 - \Gamma_{\rho}\vec{\tau} \cdot \vec{\rho}_{\mu} \: .
\end{equation}
The minimal coupling of the nucleons with mass $m$
to the mesons $i=\sigma$, $\omega$, $\delta$, $\rho$ of mass $m_{i}$
is specified by coupling strengths $\Gamma_{i}$. 
The contribution
\begin{eqnarray}
 \mathcal{L}_{m} & = & \frac{1}{2} 
 \left[ \partial^{\mu} \sigma \partial_{\mu} \sigma
  - m_{\sigma}^{2} \sigma^{2} 
+\partial^{\mu} \vec{\delta} \cdot \partial_{\mu} \vec{\delta}
  - m_{\delta}^{2} \vec{\delta} \cdot \vec{\delta}
  \right. \\ \nonumber & & \left. 
  - \frac{1}{2} G^{\mu \nu} G_{\mu \nu} 
  + m_{\omega}^{2} \omega^{\mu} \omega_{\mu} 
  - \frac{1}{2} \vec{H}^{\mu \nu} \cdot \vec{H}_{\mu \nu} 
  + m_{\rho}^{2} \vec{\rho}^{\mu} \cdot \vec{\rho}_{\mu} \right]
\end{eqnarray} 
to the total Lagrangian density $\mathcal{L}$ with
field tensors
\begin{equation}
  G_{\mu \nu}  = 
  \partial_{\mu} \omega_{\nu} - \partial_{\nu} \omega_{\mu} 
 \qquad
  \vec{H}_{\mu \nu} =
  \partial_{\mu} \vec{\rho}_{\nu} - \partial_{\nu} \vec{\rho}_{\mu} 
\end{equation}
describes the free meson fields. 
The Lagrangian density (\ref{lagdef})
is symmetrized in the covariant derivative of the
nucleon field to obtain field equations for $\psi$ and $\bar{\psi}$
that are related by a simple adjungation.

In constrast to the standard QHD Lagrangian density with 
Dirac $\gamma_{\mu}$ matrices in the kinetic term,
the $4\times 4$ matrices
\begin{equation} \label{newgam}
 \Gamma_{\mu} =  \gamma^{\nu} 
  \left[ \left(1 +W \right) g_{\mu \nu} + X_{\mu \nu}\right]   -T_{\mu} 
\end{equation}
are introduced in the DC model ($g_{\mu \nu} = \mbox{diag}
(1,-1,-1,-1)$ is the metric tensor).
They contain quantities
\begin{equation}
 W = \mathcal{W} + \vec{\tau} \cdot \vec{\mathcal{W}}
 \qquad
 T_{\mu} = \mathcal{T}_{\mu} + \vec{\tau} \cdot \vec{\mathcal{T}}_{\mu}
 \qquad
 X_{\mu \nu} = \mathcal{X}_{\mu \nu} + \vec{\tau} 
 \cdot \vec{\mathcal{X}}_{\mu \nu}
\end{equation}
which are parametrized up to 
quadratic terms in the isoscalar meson fields
\begin{equation} \label{wwdef}
 \mathcal{W} =    \frac{\Gamma_{\sigma}^{(1)}}{m} \sigma  
  - \frac{\Gamma^{(2)}_{\sigma}}{m^{4}}   m_{\sigma}^{2} \sigma^{2}
  + \frac{\Gamma^{(2a)}_{\omega}}{m^{4}} 
   m_{\omega}^{2} \omega^{\mu} \omega_{\mu}
\end{equation}
\begin{equation} \label{xxdef}
 \mathcal{T}_{\mu}  =    \frac{\Gamma_{\omega}^{(1)}}{m} \omega_{\mu}
 \qquad
 \mathcal{X}_{\mu \nu}  =    \frac{\Gamma^{(2b)}_{\omega}}{m^{4}} 
   m_{\omega}^{2} \omega_{\mu} \omega_{\nu} 
\end{equation}
and linear terms in the isovector meson fields
\begin{equation}
 \vec{\mathcal{W}}  =    \frac{\Gamma_{\delta}^{(1)}}{m} \vec{\delta}  
 \qquad
 \vec{\mathcal{T}}_{\mu}  =    \frac{\Gamma_{\rho}^{(1)}}{m} \vec{\rho}_{\mu} 
 \qquad
  \vec{\mathcal{X}}_{\mu \nu } = 0 \: .
\end{equation}
The fields $W$, $T_{\mu}$, and $X_{\mu \nu}$ describe the coupling
of the meson fields to the derivative of the nucleon field. 
The term $\mathcal{W}$ is similar to the extension of the ZM model
and leads to a density dependence of the effective self energies.
As will be seen later, the terms $\mathcal{T}_{\mu}$ 
and $\mathcal{X}_{\mu \nu}$
are the essential terms to give a momentum dependence to the
scalar and vector self energies, respectively.
Since the
covariant derivative  is used in (\ref{lagdef}), 
there appear also higher order nonlinear couplings
to the nucleon field in the Lagrangian density.

In addition to the
four coupling constants $\Gamma_{i}$ in $\mathcal{L}$
specifying the minimal nucleon-meson couplings, 
seven new couplings constants ($\Gamma_{\sigma}^{(1)}$, 
$\Gamma_{\sigma}^{(2)}$, $\Gamma_{\omega}^{(1)}$, 
$\Gamma_{\omega}^{(2a)}$, $\Gamma_{\omega}^{(2b)}$,
$\Gamma_{\delta}^{(1)}$, $\Gamma_{\rho}^{(1)}$)
appear. For the sake of convenience, mass factors are introduced
to obtain pure numbers for the new coupling constants.
The large number of new couplings makes the Lagrangian of the
DC model very flexible. 
If the new coupling constants vanish 
the matrix $\Gamma_{\mu}$ (\ref{newgam}) 
reduces to the usual $\gamma_{\mu}$
matrix and we recover the original
QHD Lagrangian density with only mininal couplings. 
In comparison to the widely employed models with
non-linear selfcouplings of the $\sigma$-meson ($\sigma$-meson
and $\omega$-meson) there are five (four) additional 
free parameters in the DC model.

The non-linear or density-dependent RMF models discussed in the
introduction are not a simple subcase of the present DC models.
They could be included by making the couplings $\Gamma_{i}$
density-dependent or replacing them by polynomials in $\sigma$
and $\omega_{\mu}$. We do not want ot make this further extension.
As will be seen later, the higher order terms in eqs.\
(\ref{wwdef}) and (\ref{xxdef}) produce similar effects as such
additional terms.

The field equations of nucleons and mesons are derived from the corresponding
Euler-Lagrange equations. In the mean field approximation the meson fields
are treated as classical fields by replacing them with their expectation
values. Only positive energy states of the nucleons are
taken into account in the calculation of the densities (no-sea
approximation).
Symmetries of the nuclear system can lead to a considerable
simplification the equations in the actual calculation, 
e.g. in infinite nuclear matter.

The field equation of the nucleons
\begin{equation} \label{feqn}
 \Gamma_{\mu} i D^{\mu} \psi - 
 \left( m - \Gamma_{\sigma}\sigma
  -\Gamma_{\delta} \vec{\tau} \cdot \vec{\delta} \right) \psi
 + \frac{1}{2} \left( i\partial^{\mu} \Gamma_{\mu}\right) \psi
 = 0
\end{equation}
can be transformed to the usual Dirac form 
\begin{equation} \label{Deq}
 \left[ \gamma^{\mu} ( i \partial_{\mu} 
 - \Sigma_{\mu})   -(m-\Sigma) \right] \psi = 0
\end{equation} 
by introducing the scalar self energy
\begin{equation} \label{ses}
 \Sigma = S - T_{\mu} i \partial^{\mu}
\end{equation}
and the vector self energy
\begin{equation} \label{sev}
 \Sigma_{\mu} = V_{\mu} - W i \partial_{\mu} 
 - X_{\mu \nu} i \partial^{\nu}
\end{equation}
with
\begin{equation}
 S =  \Gamma_{\sigma} \sigma 
  + \Gamma_{\delta} \vec{\tau} \cdot \vec{\delta}
  + T_{\mu} 
 \left( \Gamma_{\omega} \omega^{\mu}
 + \Gamma_{\rho}\vec{\tau} \cdot \vec{\rho}^{\mu} \right)
  -\frac{1}{2} (i \partial^{\mu}T_{\mu}) 
\end{equation}
and
\begin{equation}
 V_{\mu} =
  [( 1 + W )g_{\mu \nu} + X_{\mu \nu}] 
 \left( \Gamma_{\omega} \omega^{\nu}
 + \Gamma_{\rho}\vec{\tau} \cdot \vec{\rho}^{\nu}  \right) 
 -\frac{1}{2} (i \partial_{\mu} W)
 -\frac{1}{2} (i \partial^{\nu} X_{\mu \nu}) \: .
\end{equation} 
The self energies (\ref{ses}) and (\ref{sev}) in the DC model are differential
operators that act on the nucleon field. This fact is the essential
extension in the new model with derivative couplings.
Thus the self energies describe an interaction that contains 
a state dependence in addition to the medium dependence from the 
mesons fields. Additionally, both $S$ and $V_{\mu}$ depend on
scalar and vector meson fields.

From the field equation (\ref{feqn}) 
the continuity equation
$ \partial^{\mu} \hat{J}_{\mu} = 0$
for the current density operator
$ \hat{J}_{\mu} =  \bar{\psi} \Gamma_{\mu} \psi $
is immediately derived.
Therefore, $\varrho= \langle \hat{J}_{0} \rangle$ has  to be interpreted as
the conserved baryon density in the DC model. The usual vector density
$\varrho_{V} = \langle \bar{\psi} \gamma_{0} \psi \rangle$ 
is no longer a conserved
quantity. Similar we find 
$ \partial^{\mu} \hat{\vec{J}}_{\mu} = 0$
for the isospin current density operator
$ \hat{\vec{J}}_{\mu} =  \bar{\psi} \Gamma_{\mu} \vec{\tau} \psi$ .

The Lagrangian density $\mathcal{L}$ leads to the field equations
for the isoscalar mesons
\begin{eqnarray} \label{ismes}
   \partial^{\mu} \partial_{\mu} \sigma  + m_{\sigma}^{2} \sigma 
  \left(1+2\frac{\Gamma^{(2)}_{\sigma}}{m^{4}} \varrho_{s}^{D} \right)
 & = & \Gamma_{\sigma} \varrho_{s}
 + \frac{\Gamma_{\sigma}^{(1)}}{m} \varrho_{s}^{D}
  \\ \label{ivmes}
  \partial_{\mu} G^{\mu \nu} 
  + m_{\omega}^{2} \omega_{\mu}  \left[ g^{\mu \nu}
  \left(1+2\frac{\Gamma^{(2a)}_{\omega}}{m^{4}} \varrho_{s}^{D} \right) 
  +   \frac{\Gamma^{(2b)}_{\omega}}{m^{4}}
  \left( t^{D\mu \nu} + t^{D\nu \mu} \right) \right]
 & = &   \Gamma_{\omega} J^{\nu}
  +\frac{\Gamma_{\omega}^{(1)}}{m} j^{D \nu}
\end{eqnarray}
and for the isovector mesons
\begin{eqnarray} \label{mfed}
   \partial^{\mu} \partial_{\mu} \vec{\delta} 
  + m_{\delta}^{2} \vec{\delta}
 & = & \Gamma_{\delta} \vec{\varrho}_{s}
 + \frac{\Gamma_{\delta}^{(1)}}{m} \vec{\varrho}_{s}^{D}
  \\ \label{mfer}
  \partial_{\mu} \vec{H}^{\mu \nu} 
  + m_{\rho}^{2} \vec{\rho}^{\nu}  & = &
  \Gamma_{\rho} \vec{J}^{\nu}
  +\frac{\Gamma_{\rho}^{(1)}}{m} \vec{\jmath}^{D \nu} \: .
\end{eqnarray}
They contain additional terms as compared to
the linear QHD model.
The masses of the isoscalar mesons are no longer constant
but become effectively density dependent.
The minimal coupling of nucleons and
mesons leads to source terms with the usual scalar densities
\begin{equation}
 \varrho_{s} = \langle \bar{\psi} \psi \rangle
 \qquad
 \vec{\varrho}_{s} = \langle \bar{\psi} \vec{\tau} \psi \rangle \: ,
\end{equation}
and the new current densities
\begin{equation}
 J_{\mu} = \langle \bar{\psi} \Gamma_{\mu} \psi \rangle
 \qquad
 \vec{J}_{\mu} = \langle \bar{\psi} \Gamma_{\mu} \vec{\tau} \psi \rangle 
\end{equation}
which depend on
the usual vector current densities
\begin{equation} 
 j_{\mu} = \langle \bar{\psi} \gamma_{\mu} \psi \rangle
 \qquad 
 \vec{\jmath}_{\mu} = \langle \bar{\psi} \gamma_{\mu} \vec{\tau} \psi \rangle 
\end{equation}
and the scalar densities (see appendix \ref{appa}).

Additional source terms proportional
to $\Gamma_{\sigma}^{(1)}$, $\Gamma_{\omega}^{(1)}$, $\Gamma_{\delta}^{(1)}$,
and $\Gamma_{\rho}^{(1)}$  
appear in the field equations with the
derivative scalar densities
\begin{eqnarray} \label{rhosD1}
 \varrho_{s}^{D} & = & 
 \frac{1}{2} \langle \left[ \bar{\psi} \gamma^{\mu} i D_{\mu} \psi 
 + \overline{(iD_{\mu}\psi)} \gamma^{\mu} \psi \right] \rangle
 = g^{\mu \nu} t_{\mu \nu}^{D}
 \\  \label{rhosD2}
  \vec{\varrho}_{s}^{D} & = & 
 \frac{1}{2} \langle \left[ \bar{\psi} \gamma^{\mu} i D_{\mu} \vec{\tau} \psi 
 + \overline{(iD_{\mu}\psi)} \gamma^{\mu} \vec{\tau} \psi \right] \rangle
 = g^{\mu \nu} \vec{t}_{\mu \nu}^{D}
\end{eqnarray}
and the derivative current densities
\begin{eqnarray}
 j^{D}_{\mu} & = &
 \frac{1}{2} \langle \left[ \bar{\psi}  i D_{\mu} \psi 
 + \overline{(iD_{\mu}\psi)}  \psi \right] \rangle
 \\ 
 \vec{\jmath}^{D}_{\mu} & = &
 \frac{1}{2} \langle \left[ \bar{\psi}  i D_{\mu} \vec{\tau}\psi 
 + \overline{(iD_{\mu}\psi)}  \vec{\tau} \psi \right] \rangle \: .
\end{eqnarray}
As shown, the scalar densities (\ref{rhosD1}) and (\ref{rhosD2})
can also be expressed in terms of the derivative tensor densities
\begin{eqnarray}
 t^{D}_{\mu \nu} & = &
 \frac{1}{2} \langle \left[ \bar{\psi} \gamma_{\mu} i D_{\nu} \psi 
 + \overline{(iD_{\nu}\psi)} \gamma_{\mu} \psi \right] \rangle
 \\
  \vec{t}^{D}_{\mu \nu} & = &
 \frac{1}{2} \langle \left[ \bar{\psi} \gamma_{\mu} i D_{\nu} \vec{\tau} \psi 
 + \overline{(iD_{\nu}\psi)} \gamma_{\mu} \vec{\tau} \psi \right] \rangle
\end{eqnarray}
which are not symmetric in the
Lorentz indices. More explicit expressions for the densities can be found in
appendix \ref{appa}.

\section{Application to infinite nuclear matter}

Infinite nuclear matter in its ground state is a homogeneous, 
isotropic and stationary system. Because of these symmetries, 
the set of field equations simplifies considerably.
Densities do not depend on space-time coordinates and only the
time-like component of the currents remains. Correspondingly,
the meson fields are constant and the spatial components of the
vector meson field vanish. Additionally,
only the third component in isospin of the isovector densities and isovector
meson fields remains. The matrix $\Gamma_{\mu}$ is diagonal in isospin space
and it is useful to introduce the abbreviations
\begin{equation}
 \Gamma_{0 \pm} =  \gamma_{0} 
  \left(1 +W_{\pm} + X_{00 \pm} \right)- T_{0 \pm} 
 \qquad
 \Gamma_{i \pm} = \gamma_{i} \left(1+W_{\pm}\right) \quad i=1,2,3
\end{equation}
for the time-like and space-like components of $\Gamma_{\mu}$
with the quantities
\begin{equation}
 W_{\pm} = \mathcal{W} \pm \left(\vec{\mathcal{W}}\right)_{3}
 \qquad
 T_{0 \pm} = \mathcal{T}_{0} \pm \left(\vec{\mathcal{T}}_{0}\right)_{3}
 \qquad
 X_{00 \pm} = \mathcal{X}_{00} \pm \left(\vec{\mathcal{X}}_{00}\right)_{3}
\end{equation}
where the case of protons is denoted by  $+$ and neutrons by $-$,
respectively.
The momentum independent contributions to the
self energies (\ref{ses}) and (\ref{sev}) are given by
\begin{equation}
 S_{\pm} =  \Gamma_{\sigma} \sigma 
  \pm \Gamma_{\delta} (\vec{\delta})_{3}
  + T_{0 \pm} 
 \left[ \Gamma_{\omega} \omega_{0}
 \pm \Gamma_{\rho} (\vec{\rho}_{0})_{3} \right]
\end{equation}
and
\begin{equation}
 V_{0 \pm} =
  ( 1 + W_{\pm}  + X_{00 \pm})
 \left[ \Gamma_{\omega} \omega_{0}
 \pm \Gamma_{\rho} \left(\vec{\rho}_{0}\right)_{3}  \right] \: .
\end{equation} 
The meson fields are directly obtained from the
field equations (\ref{ismes}) - (\ref{mfer}) for given densities.

\subsection{Solutions of the generalized Dirac equation}

In order to calculate the
various densities appearing in the field equations we first have
to find the nucleon field from the the generalized Dirac equation 
with self-energies
that are differential operators in the DC model.  
Solutions of the nucleon field equation are found 
with the plane wave ansatz
\begin{equation} \label{ansatz}
 \psi(\vec{p},\sigma, \tau) 
 = u(\vec{p},\sigma, \tau) \exp \left( -i p_{\mu} x^{\mu} \right)
\end{equation}
for a nucleon with four momentum $p_{\mu} = ( E , \vec{p})$ 
and positive energy
$E$. Spin and isospin quantum numbers are
specified by $\sigma$ and $\tau$.
The scalar self energy
\begin{equation}
 \Sigma_{\pm} = S_{\pm} - T_{0\pm} E
\end{equation}
and the vector self energy
\begin{equation}
 \Sigma_{0\pm} = V_{0\pm} - (W_{\pm} +  X_{00\pm}) E
\end{equation}
for nucleons in nuclear matter
are explicitly energy dependent in the DC model. They decrease with energy
for positive $T_{0\pm}$ and $(W_{\pm}+X_{00\pm})$. Of course, this linear
dependence will give a reasonable description of realistic self energies
only for not too high energies. The quantities $T_{0\pm}$, $W_{\pm}$,
and $X_{00\pm}$ increase with increasing density and the energy dependence
of the self energies becomes stronger. 
This behaviour also leads to a limitation in the density range
where the DC model can be applied because of the increase of the
Fermi momentum.
Since the energy $E$ already contains the
rest mass $m$ of the nucleon the self energies are rather insensitive to the
nucleon energy if $E<m$.
The energy dependence itself is density dependent and it
is different for protons and neutrons if $\Gamma_{\delta}^{(1)}\neq 0$ or
$\Gamma_{\rho}^{(1)}\neq 0$.
The ansatz (\ref{ansatz}) in the Dirac equation (\ref{Deq}) 
leads to the condition
\begin{equation} \label{Deqp}
 \left( \begin{array}{cc}
 \tilde{M}_{\tau} - \tilde{E}_{\tau} 
 & \vec{\sigma} \cdot \vec{\tilde{p}}_{\tau}  \\
 \vec{\sigma} \cdot \vec{\tilde{p}}_{\tau}  
 & -\tilde{M}_{\tau} - \tilde{E}_{\tau}
 \end{array}\right)
 u( \vec{p},\sigma, \tau )  = 0
\end{equation}
for the spinor $u( \vec{p},\sigma, \tau )$ if the abbreviations
\begin{eqnarray} \label{mtt}
 \tilde{M}_{\tau} & = & m - S_{\tau} + T_{0\tau} E
 \\ \label{ett}
 \tilde{E}_{\tau} & = & (1+W_{\tau}+X_{00 \tau}) E - V_{0 \tau} 
 \\ \label{ptilde}
 \vec{\tilde{p}}_{\tau} & = & (1+W_{\tau}) \vec{p}
\end{eqnarray}
are introduced. These quantities 
carry the index $\tau = \pm $ because the fields
are not necessarily equal for
protons and neutrons in asymmetric 
nuclear matter. The dispersion relation
\begin{equation} \label{disp}
 \tilde{M}_{\tau}^{2} = 
 \tilde{E}_{\tau}^{2}  - \tilde{p}_{\tau}^{2} 
\end{equation}
connects the energy 
\begin{equation} 
  \tilde{E}_{\tau}  = \frac{1}{\sqrt{1-B^{2}}} \left( A_{\tau} B_{\tau}
 + \sqrt{A_{\tau}^{2}+\tilde{p}_{\tau}^{2}} \right)
\end{equation}
and mass
\begin{equation}
 \tilde{M}_{\tau}  =  \frac{1}{\sqrt{1-B_{\tau}^{2}}} 
 \left( A_{\tau} 
 + B_{\tau} \sqrt{A_{\tau}^{2}+\tilde{p}_{\tau}^{2}} \right) \: .
\end{equation}
where the constants
\begin{equation}
 A_{\tau} = \frac{m - S_{\tau} + B_{\tau}V_{0 \tau}}{\sqrt{1-B_{\tau}^{2}}}
\qquad 
 B_{\tau} = \frac{T_{0 \tau}}{1+W_{\tau}+X_{00 \tau}} 
\end{equation}
have been introduced. 
Not only the energy $\tilde{E}_{\tau}$, 
but also the mass $\tilde{M}_{\tau}$, depends on the momentum,
the density and isospin.
Obviously, reasonable solutions of the Dirac equation
exist only for $-1<B_{\tau}<1$
and $1+W_{\tau}+X_{00 \tau}\neq 0$.
The energy in the four-momentum of the nucleon in (\ref{ansatz}) is given by
\begin{equation} \label{etau}
 E_{\tau} = (1+W_{\tau}+X_{00 \tau})^{-1}  \left[ V_{0\tau}
 +\frac{1}{\sqrt{1-B_{\tau}^{2}}} \left( A_{\tau}B_{\tau}
 + \sqrt{A_{\tau}^{2}+\tilde{p}_{\tau}^{2}} \right)  \right] \: .
\end{equation}
Solutions of equation (\ref{Deqp}) are found to be
\begin{equation} \label{spinsol}
 u ( \vec{p}, \sigma, \tau )
  = \sqrt{\frac{\tilde{M}_{\tau}+\tilde{E}_{\tau}}{2[
 (1+W_{\tau}+X_{00 \tau})\tilde{E}_{\tau}-T_{0 \tau}\tilde{M}_{\tau}]}}
 \left( \begin{array}{c} \chi_{\sigma} \\ 
 \frac{\vec{\sigma}\cdot\vec{\tilde{p}}_{\tau}}{\tilde{M}_{\tau}
 +\tilde{E}_{\tau}}\chi_{\sigma}
\end{array} \right) \xi_{\tau}
\end{equation}
with spin and isospin eigenfunctions $\chi_{\sigma}$ and $\xi_{\tau}$.
The normalization of these spinors is discussed in appendix \ref{appb}.

Negative energy solutions are obtained by a similar procedure..
Then the theory can be quantized in the usual way by expanding
the nucleon field operator in terms of the complete set of solutions
and imposing anticommutation relations for the creation and
annihilation operators. However, in the mean field solution 
we do not need to carry out this standard
procedure in detail here.

\subsection{Densities}

The densities that appear in the field equations are obtained by a
summation over all occupied states in nuclear matter up to
the Fermi momentum $p^{F}_{\tau}$ for protons and neutrons, respectively. 
In the no-sea approximation only states of positive energy are considered.
Proton and neutron constribution to the various densities can be obtained
with the help of the projection operator 
$P_{\tau} = (1+\tau (\vec{\tau})_{3})/2$,
e.g., the total baryon density
\begin{equation}
 \varrho = J_{0} = \varrho_{+} + \varrho_{-}
\end{equation}
and the third component of the isospin density
\begin{equation}
 \left(\vec{J}_{0}\right)_{3} = \varrho_{+} - \varrho_{-}
\end{equation}
are the sum and the difference of 
the proton and neutron densities
\begin{equation}
 \varrho_{\tau} = \langle \bar{\psi} \Gamma_{0} P_{\tau} \psi \rangle \: ,
\end{equation}
respectively.
The other isoscalar and isovector densities are related to the corresponding
proton and neutron densities in a similar manner.
Considering the normalization of the solutions of the Dirac equation
all densities can be  expressed in terms of
the three fundamental integrals
\begin{eqnarray}
 I^{\tau}_{1} & = & \int_{0}^{\tilde{p}^{F}_{\tau}} d\tilde{p}_{\tau} \: 
 \tilde{p}_{\tau}^{2}
 = \frac{1}{3} (\tilde{p}^{F}_{\tau})^{3}
 \\
 I^{\tau}_{2} & = & \int_{0}^{\tilde{p}^{F}_{\tau}} d\tilde{p}_{\tau} \: 
 \frac{\tilde{p}_{\tau}^{2}}{\sqrt{A_{\tau}^{2}+\tilde{p}_{\tau}^{2}}}
 = \frac{1}{2} \left[ \tilde{p}^{F}_{\tau} E^{F}_{\tau} -
 A_{\tau}^{2} \ln \frac{\tilde{p}^{F}_{\tau}
 +E^{F}_{\tau}}{A_{\tau}}\right]
  \\
 I^{\tau}_{3} & = & \int_{0}^{\tilde{p}^{F}_{\tau}} d\tilde{p}_{\tau} \: 
 \tilde{p}_{\tau}^{2} \sqrt{A_{\tau}^{2}+\tilde{p}_{\tau}^{2}} = 
 \frac{3E^{F}_{\tau}}{4} I^{\tau}_{1} 
 + \frac{A_{\tau}^{2}}{4} I^{\tau}_{2} 
\end{eqnarray}
with
$ \tilde{p}^{F}_{\tau} = (1+W) p^{F}_{\tau}$
and
$ E^{F}_{\tau} = \sqrt{A_{\tau}^{2}+(\tilde{p}^{F}_{\tau})^{2}}$.
The proton and neutron densities are then given by
\begin{equation}
 \varrho_{\tau} 
 = \langle \bar{\psi} \Gamma_{0} P_{\tau} \psi \rangle
 = \frac{\kappa}{2\pi^{2}} (1+W)^{-3} I^{\tau}_{1}
 = \frac{1}{3\pi^{2}} \left( p^{F}_{\tau} \right)^{3}
\end{equation}
with the spin degeneracy factor $\kappa=2$. This relation defines the
corresponding Fermi momentum $p^{F}_{\tau}$ for a given nucleon density.
Immediately, we obtain the scalar density
\begin{equation}
  \varrho_{s\tau} 
 = \langle \bar{\psi} P_{\tau} \psi \rangle 
 = C_{\tau}^{-1} 
 \left[ B_{\tau}I^{\tau}_{1} + A_{\tau} I^{\tau}_{2} \right]
\end{equation}
and the current density
\begin{equation}
 j_{0\tau} 
 = \langle \bar{\psi} \gamma_{0} P_{\tau} \psi \rangle
 = C_{\tau}^{-1} 
 \left[ I^{\tau}_{1} + A_{\tau} B_{\tau} I^{\tau}_{2} \right] 
\end{equation}
with
\begin{equation}
 C_{\tau} = \pi^{2} (1+W_{\tau}+X_{00 \tau})  (1+W_{\tau})^{3} 
 (1-B_{\tau}^{2}) \: .
\end{equation}
With these results the relation
\begin{equation}
 \varrho_{\tau} = (1+W_{\tau}+X_{00 \tau}) j_{0\tau} - T_{0\tau} 
 \varrho_{s\tau} 
\end{equation}
is easily confirmed.
Considering the relation (\ref{ett})  the 
derivative densities are most easily calculated from
\begin{equation}
 j^{d}_{0\tau} = 
 \langle \bar{\psi} E_{\tau} P_{\tau} \psi \rangle
 =  (1+W_{\tau}+X_{00\tau})^{-1}   \left( \varrho^{E}_{s\tau}  +  
 V_{0\tau} \varrho_{s\tau}
 \right)
\end{equation}
and
\begin{equation}
 t^{d}_{00\tau} = 
 \langle \bar{\psi} \gamma_{0} E_{\tau} P_{\tau} \psi \rangle
 =  (1+W_{\tau}+X_{00\tau})^{-1}   \left( \varrho^{E}_{\tau}  +  
 V_{0\tau} j_{0\tau} \right)
\end{equation}
where
\begin{equation}
 \varrho^{E}_{s\tau} =  
 \langle \bar{\psi} \tilde{E}_{\tau} P_{\tau} \psi \rangle
  =   C_{\tau}^{-1}(1-B_{\tau}^{2})^{-\frac{1}{2}}
 \left[  A_{\tau}(1+B_{\tau}^{2}) I^{\tau}_{1} 
 + A_{\tau}^{2}B_{\tau} I^{\tau}_{2}
  + B_{\tau} I^{\tau}_{3}  \right]
\end{equation}
and 
\begin{equation}
 \varrho^{E}_{\tau} =  
 \langle \bar{\psi} \gamma_{0} \tilde{E}^{+}_{\tau} P_{\tau} \psi \rangle
  =      C_{\tau}^{-1}(1-B_{\tau}^{2})^{-\frac{1}{2}}
 \left[   2 A_{\tau}B_{\tau} I^{\tau}_{1} 
 + A_{\tau}^{2}B_{\tau}^{2} I^{\tau}_{2}
  + I^{\tau}_{3}  \right] \: .
\end{equation}
The scalar derivative density is deduced from
\begin{eqnarray}
 \varrho_{s \tau}^{d} & = & t_{\lambda \mu _{\tau}}^{d} g^{\lambda \mu}
 =   \langle \bar{\psi} \gamma_{0} E_{\tau} P_{\tau} \psi \rangle
 - \langle \bar{\psi} \vec{\gamma} \cdot \vec{p} P_{\tau} \psi \rangle
 \\ \nonumber & = &
 t_{00\tau}^{d} -   (1+W_{\tau})^{-1} 
 \left( \varrho^{E}_{\tau} - \varrho^{M}_{s\tau} \right)
\end{eqnarray}
with (\ref{ugpu}), the dispersion relation (\ref{disp}), and
\begin{equation}
 \varrho^{M}_{s\tau} =
  \langle \bar{\psi} \tilde{M}_{\tau}  P_{\tau} \psi \rangle
  =  C_{\tau}^{-1}(1-B_{\tau}^{2})^{-\frac{1}{2}}
  \left[ 2 A_{\tau} B_{\tau} I^{\tau}_{1} + A_{\tau}^{2} I^{\tau}_{2}
  + B_{\tau}^{2} I^{\tau}_{3} \right] \: .
\end{equation}

Nuclear matter is characterized by its equation of state,
i.e. the energy per nucleon or pressure as a function of 
the nucleon density. The energy density  
is calculated from the energy-momentum
tensor
\begin{eqnarray}
 T^{\lambda \mu} & = & \sum_{\varphi}
 \frac{\partial \mathcal{L}}{\partial (\partial_{\lambda} \varphi )}
 \partial^{\mu} \varphi - g^{\lambda \mu} \mathcal{L}
\end{eqnarray}
where the sum runs over all fields ($\varphi= \psi,\bar{\psi},\sigma,
 \vec{\delta},\omega_{\nu},\vec{\rho}_{\nu}$).
Since there are no contributions from the derivatives
of the meson fields in nuclear matter the expectation value of the
energy-momentum tensor reduces to
\begin{equation}
 \langle T^{\lambda \mu} \rangle 
  =    \langle \bar{\psi}  \Gamma^{\lambda} p^{\mu}
 \psi \rangle- g^{\lambda \mu} \langle \mathcal{L} \rangle 
\end{equation}
with
\begin{eqnarray}
 \langle \mathcal{L} \rangle =
 \frac{1}{2}  \left[  m_{\omega}^{2} \omega^{\mu} \omega_{\mu} 
  + m_{\rho}^{2} \vec{\rho}^{\mu} \cdot \vec{\rho}_{\mu} 
  - m_{\sigma}^{2} \sigma^{2} 
  - m_{\delta}^{2} \vec{\delta} \cdot \vec{\delta}  \right] \: .
\end{eqnarray} 
Only the meson fields contribute to $\langle \mathcal{L} \rangle$
because the nucleon fields are solutions of the Dirac equation (\ref{feqn}).
With the help of equations (\ref{mtt}), (\ref{ett}), and
(\ref{ugpu}) the energy density 
\begin{eqnarray}
 \varepsilon & = & \langle T_{00} \rangle
 =  \langle \bar{\psi} \Gamma_{0} E_{0} \psi \rangle
 - \langle \mathcal{L} \rangle 
 \\ \nonumber & = &  \sum_{\tau} \langle   \bar{\psi} \left[
 \gamma_{0} (\tilde{E}_{\tau} +V_{0\tau})
 - (\tilde{M}_{\tau} -m + S_{\tau}) \right] P_{\tau} \psi  \rangle
 - \langle \mathcal{L} \rangle 
 \\ \nonumber & = &  \sum_{\tau} \left[ \varrho^{E}_{\tau}
 - \varrho^{M}_{s\tau}  + (m - S_{\tau}) \varrho_{s\tau} 
  + V_{0\tau} j_{0\tau} 
 \right]
 \\ \nonumber & & 
 - \frac{1}{2}  \left[  m_{\omega}^{2} \omega_{0}^{2}
  + m_{\rho}^{2} \vec{\rho}_{0}^{2} 
  - m_{\sigma}^{2} \sigma^{2} 
  - m_{\delta}^{2} \vec{\delta}^{2}  \right] 
\end{eqnarray}
and the pressure 
\begin{eqnarray} \label{press}
 p & = & \frac{1}{3} \sum_{i=1}^{3} \langle T_{ii} \rangle
 = \frac{1}{3} \langle   \bar{\psi} \vec{\gamma} 
 \cdot \vec{\tilde{p}} \psi  \rangle  + \langle \mathcal{L} \rangle 
 \\ \nonumber & = & 
 \frac{1}{3} \sum_{\tau} \left( \varrho^{E}_{\tau}
  - \varrho^{M}_{s\tau} \right)
 + \frac{1}{2}  \left[  m_{\omega}^{2} \omega_{0}^{2}
  + m_{\rho}^{2} \vec{\rho}_{0}^{2}
  - m_{\sigma}^{2} \sigma^{2} 
  - m_{\delta}^{2} \vec{\delta}^{2}  \right]   
\end{eqnarray}
are found.
The DC model in the mean field approximation is
thermodynamically consistent because the pressure (\ref{press})
calculated from the energy-momentum tensor agrees with
the thermodynamical pressure $p_{th} = \varrho^{2} \frac{d}{d\varrho}
\left(\frac{\varepsilon}{\varrho}\right)$.
Furthermore the relation 
\begin{equation}
 \varepsilon + p =  \mu_{+} \varrho_{+} +  \mu_{-} \varrho_{-} 
\end{equation}
is obtained with the chemical potentials
\begin{equation} 
 \mu_{\tau} = (1+W_{\tau}+X_{00 \tau})^{-1}  \left[ V_{0\tau}
 +\frac{1}{\sqrt{1-B_{\tau}^{2}}} \left( A_{\tau}B_{\tau}
 + \sqrt{A_{\tau}^{2}+(\tilde{p}_{\tau}^{F})^{2}} \right)  \right] 
\end{equation}
which are the energies (\ref{etau}) of the nucleons at the Fermi surface
and the Hugenholtz-van Hove theorem \cite{Hug58} holds.

Determing the properties of nuclear matter in the DC model for a given
density $\varrho$ and neutron-proton asymmetry
$\alpha= \frac{\varrho_{n}-\varrho_{p}}{\varrho_{n}+\varrho_{p}}$
is more involved than in standard RMF models. The meson fields have 
to be determined from the densities which themselves 
depend on the meson fields.
A self-consistent solution is achieved by iteratively solving the set of
equations until convergence is reached.

\subsection{Optical potential}

The Schr\"{o}dinger equivalent optical potential $V_{\rm opt}$ serves as a 
convenient means to characterize the in-medium interaction of a
nucleon. It is obtained by recasting the Dirac equation (\ref{Deq})
into a Schr\"{o}dinger-like equation for the large (upper) component of the
nucleon spinor. In the DC model the momentum $\vec{p}$ in the 
generalized Dirac equation is multiplied by a factor $(1+W_{\tau})$
in eq.\ (\ref{ptilde}).
In order to start for the non-relativistic reduction
from the usual form of the time-independent Dirac equation 
\begin{equation} \label{Deq2}
 \left[ \vec{\alpha} \cdot \vec{p} + 
 \beta \left( m -\Sigma^{\rm eff}_{\tau} \right)
 \right] \psi_{\tau}
 = \left( E - \Sigma_{0 \tau}^{\rm eff} \right) \psi_{\tau}
\end{equation}
the effective self energies
\begin{equation} \label{sseff}
 \Sigma^{\rm eff}_{\tau}  =  \frac{\Sigma_{\tau}+W_{\tau}m}{1+W_{\tau}}
 = \frac{S_{\tau}-T_{0\tau}E+W_{\tau}m}{1+W_{\tau}}
\end{equation} 
and
\begin{equation} \label{sveff}
 \Sigma_{0\tau}^{\rm eff}  =  \frac{\Sigma_{0\tau}+W_{\tau}E}{1+W_{\tau}}
 = \frac{V_{0\tau}-X_{00\tau}E}{1+W_{\tau}}
\end{equation}
are introduced. They depend linearly  on the nucleon energy
$E$ if $T_{0\tau} \neq 0$ and $X_{00\tau} \neq 0$.
Since $E$ contains the rest mass of the nucleon, there is only
a small variation of the self-energies at small momenta of the nucleon.
It is seen that the coupling vertices in the DC model with 
$\Gamma_{\omega}^{(1)}$,
$\Gamma_{\rho}^{(1)}$,
and $\Gamma_{\omega}^{(2b)}$ lead to the energy-dependent
scalar and vector self-energies, respectively. 
In the DC model the energy dependence
results from the coupling of the Lorentz-vector meson fields
to the derivative of the nucleon field. The Lorentz-scalar meson
fields do not lead to an energy dependence of the effective self energies.
The modification of the initial Lagrangian (\ref{lagdef}) by the term $W$ 
alone does not introduce an energy dependence of the effective self energies.
This resembles the ZM model in which the derivative coupling 
is a special case of the $W$ contribution in the DC model. 
By a rescaling of the nucleon wave function in the ZM model, 
the derivative coupling 
was essentially removed, leading to a nonlinear coupling of the nucleon
field to the $\sigma$-meson without an energy dependence of the self energies.

The quantities $T_{0\tau}$ and $X_{00\tau}$ have  different
density dependencies. At low densities they are approximately
proportional to the baryon density and the square of the baryon
density, respectively. But with increasing density this proportionality is lost
because of the nonlinearity of the meson field equations.
Even without an energy dependence the effective self energies
(\ref{sseff}) and (\ref{sveff}) in the DC model show a nontrivial 
density dependence.

There are various methods to derive the optical potential from
the Dirac equation (\ref{Deq2}). Here, the definition
\begin{equation} \label{voptdef}
 V_{\rm opt \tau} = \frac{E}{m} \Sigma_{0\tau}^{\rm eff} 
- \Sigma^{\rm eff}_{\tau}
 + \frac{1}{2m} \left[ \left( \Sigma^{\rm eff}_{\tau} \right)^{2}
 - \left( \Sigma_{0\tau}^{\rm eff} \right)^{2} \right]
\end{equation}
for the Schr\"{o}dinger equivalent optical potential is used
as in Refs.\ \cite{Lee97}. It has the property that $V_{\rm opt \tau}$ 
rises linearly with the
energy $E$ of the nucleon if the self energies are energy independent.
An alternative choice is the definition as used in Ref.\ \cite{Ham90} which 
corresponds to multiplying (\ref{voptdef}) by $m/E$. In Ref.\ \cite{Fel91,Li93}
the optical potential is calculated from the difference of the energy
$E$ of the nucleon in the medium and the kinetic energy of a free nucleon 
without interaction at the same momentum $\vec{p}$. This definition
is more appropriate for a general relativistic definition 
of an optical potential
but it is no longer a Schr\"{o}dinger equivalent potential.
It always leads to a constant optical potential
for large nucleon momenta for energy independent self energies.
For a meaningful comparison of optical potentials it is, of course,
necessary to chose the same definition.

The optical potential (\ref{voptdef}) in the DC model is a quadratic
function of the nucleon energy because the effective self energies 
(\ref{sseff}) and (\ref{sveff}) are linear functions of the energy.
This behaviour is reasonable
only in a limited range of energies and the model cannot be applied
at arbitrarily high energies. But for nucleon energies
up to about 1~GeV the linear dependence is a sufficient approximation. 
Indeed, self energies from Dirac
phenomenology of proton scattering show an approximately linear
decrease for small energies \cite{Ham90,Coo93}.

\section{Parametrizations and properties of nuclear matter}
\label{Sec4}

The DC model introduces a large number of additional couplings as compared 
to the original RMF model with minimal meson-nucleon couplings
and even with respect to usual non-linear models.
The corresponding coupling constants have to be fixed by the limited
number of nuclear matter properties. 
Here, we develop parametrizations within restricted versions of the DC model
in order to investigate whether the model gives reasonable results.

Symmetric nuclear matter is essentially characterized by the
binding energy per nucleon 
$E/A$ 
and the incompressibility $K$
at saturation density $\varrho_{\rm sat}$ where the pressure $p$ is zero.
This leads to three conditions for the seven isoscalar coupling constants
$\Gamma_{\sigma}$, $\Gamma_{\sigma}^{(1)}$, 
$\Gamma_{\sigma}^{(2)}$, $\Gamma_{\omega}$ $\Gamma_{\omega}^{(1)}$, 
$\Gamma_{\omega}^{(2a)}$, and $\Gamma_{\omega}^{(2b)}$. Furthermore,
an adjustment of the relativistic effective mass
$ M^{\rm eff}  = m-\Sigma^{\rm eff}$
at saturation density can be used to achieve a reasonable spin-orbit splitting
in finite nuclei. A reduction of $M^{\rm eff}$ increases the spin-orbit
splitting. The four isovector coupling constants
$\Gamma_{\delta}$, $\Gamma_{\delta}^{(1)}$, $\Gamma_{\rho}$, and 
$\Gamma_{\rho}^{(1)}$ in the DC model determine the
symmetry energy $J$ and its derivative with respect to
the density which can be parametrized by $L= 3 \varrho\frac{dJ}{d\varrho}$.
In all our parametrizations, the $\delta$-meson is not considered,
which is common practice in most RMF parametrizations, because
its contribution seems not to be well determined in parameter fits to
properties of finite nuclei.

All our parametrizations of the DC models are fitted to give $E/A=-16.0$~MeV
for the binding energy per nucleon at a saturation density of 
$\varrho_{\rm sat} = 0.150$~fm${}^{-3}$ with an incompressibility of
$K=240$~MeV. These are reasonable values as compared to the properties
of symmetric nuclear matter in other models. The parameters are
adjusted to an effective nucleon mass at $\varrho_{\rm sat}$
of $M^{\rm eff} = 0.6 m$
close to the value of several relativistic mean field parametrizations
with spin-orbit splittings of single-particle levels 
close to the experimental values.
These four conditions require at least four isoscalar coupling constants
in the model. The assumemd symmetry energy of $J=32$~MeV fixes one
isovector coupling constant.

In the first parametrization (DC1) only the usual
minimal meson-nucleon couplings with $\Gamma_{\sigma}$, 
$\Gamma_{\omega}$, and $\Gamma_{\rho}$ 
and the new couplings $\Gamma_{\sigma}^{(2)}$ and
$\Gamma_{\omega}^{(2a)}$ are assumed to be non-zero. This leads to
a model without energy-dependent effective self-energies because
$T_{\mu}=0$ and $X_{\mu \nu}=0$. The non-zero quantity $W$ contains only
contributions quadratic in the $\sigma$-meson and $\omega$-meson fields
and correspondingly, these mesons acquire a density-dependent effective mass. 
The DC1 parametrization has the same number of parameters as the
standard non-linear models with self-interactions of the $\sigma$-meson.

The second and third parameter set (DC2/DC3) are chosen 
to give a reasonable description
of the Schr\"{o}dinger equivalent optical potential of the nucleon in 
symmetric nuclear matter at saturation density $\varrho_{\rm sat}$.
Since the optical potential extracted 
in Dirac phenomenology from proton-nucleus scattering becomes almost
constant around the kinetic energy $E-m=1$~GeV of the nucleon, 
we adjust the coupling constants to exhibit a
maximum of the potential at this energy. The DC model is not really
applicable to higher energies because 
the empirical self-energies show 
a deviation from the linear energy-dependence and in our approach
the optical potential is a quadratic function of the nucleon energy.
The coupling constants $\Gamma_{\omega}^{(1)}$
and $\Gamma_{\omega}^{(2b)}$ have to be positive to generate a reduction
of the self-energies with increasing nucleon energy. In order to keep
the number of parameters small we require in this model
the additional constraints
$\Gamma_{\omega}^{(2a)}+\Gamma_{\omega}^{(2b)}=0$ and
$\Gamma_{\sigma}^{(1)}=\Gamma_{\omega}^{(1)}$. 
For small densities this condition cancels the density dependence
of the $\omega$-meson mass in eq.~(\ref{ivmes}) and makes the 
coupling coefficients to the derivative densities 
equal in eqs.~(\ref{ismes},\ref{ivmes}).
Now the 
five independent isoscalar parameters are uniquely determined
by five conditions.

In the parametrizations DC1 and DC2 the constant
$\Gamma_{\rho}$ is fitted to give a symmetry energy of $J=32$~MeV
wheras $\Gamma_{\rho}^{(1)}$ is assumed to be zero. 
The sets DC2 and DC3 share the same isoscalar coupling constants
but in model DC3 the two coupling constants of the $\rho$-meson
are adjusted to give both a symmetry energy of $J=32$~MeV and 
a derivative $L = 40$~MeV of the symmetry energy at saturation density
in symmetric nuclear matter. This value of $L$
is typical for non-relativistic Skyrme Hartree-Fock models.
An increase of $\Gamma_{\rho}^{(1)}$ leads to a decrease of 
$\Gamma_{\rho}$ in order to give the same symmetry energy.
Precise values for the coupling constants in all DC models
are presented in Table \ref{tab1}. 

\begin{table}[t]
\caption{\label{tab1} Coupling constants in the DC models. The nucleon
mass is $m=939$~MeV and the meson masses
are $m_{\sigma}=550$~MeV, $m_{\omega}=783$~MeV, and $m_{\rho}=763$~MeV.}
\begin{tabular}{cccccccccc}
 \hline \hline
 model &
 $\Gamma_{\sigma}$ & $\Gamma_{\omega}$ & $\Gamma_{\rho}$ &
 $\Gamma_{\sigma}^{(1)}$  & $\Gamma_{\omega}^{(1)}$ & $\Gamma_{\rho}^{(1)}$ &
 $\Gamma_{\sigma}^{(2)}$ & $\Gamma_{\omega}^{(2a)}$ &
 $\Gamma_{\omega}^{(2b)}$ \\
 \hline
 DC1 & 10.996338 & 12.831929 & 3.783530 & 0.000 & 0.000 & 0.000 
 & 84.77 & 12.30 & 0.00 \\
 DC2 & 10.536751 & 13.754212 & 3.917768 & 2.392 & 2.392 & 0.000 
 & 8.735 & -211.54 & 211.54 \\
 DC3 & 10.536751 & 13.754212 & 0.943337 & 2.392 & 2.392 & 5.525 
 & 8.735 & -211.54 & 211.54 \\
 \hline \hline
\end{tabular}
\end{table}

\begin{figure}
\includegraphics[width=13cm]{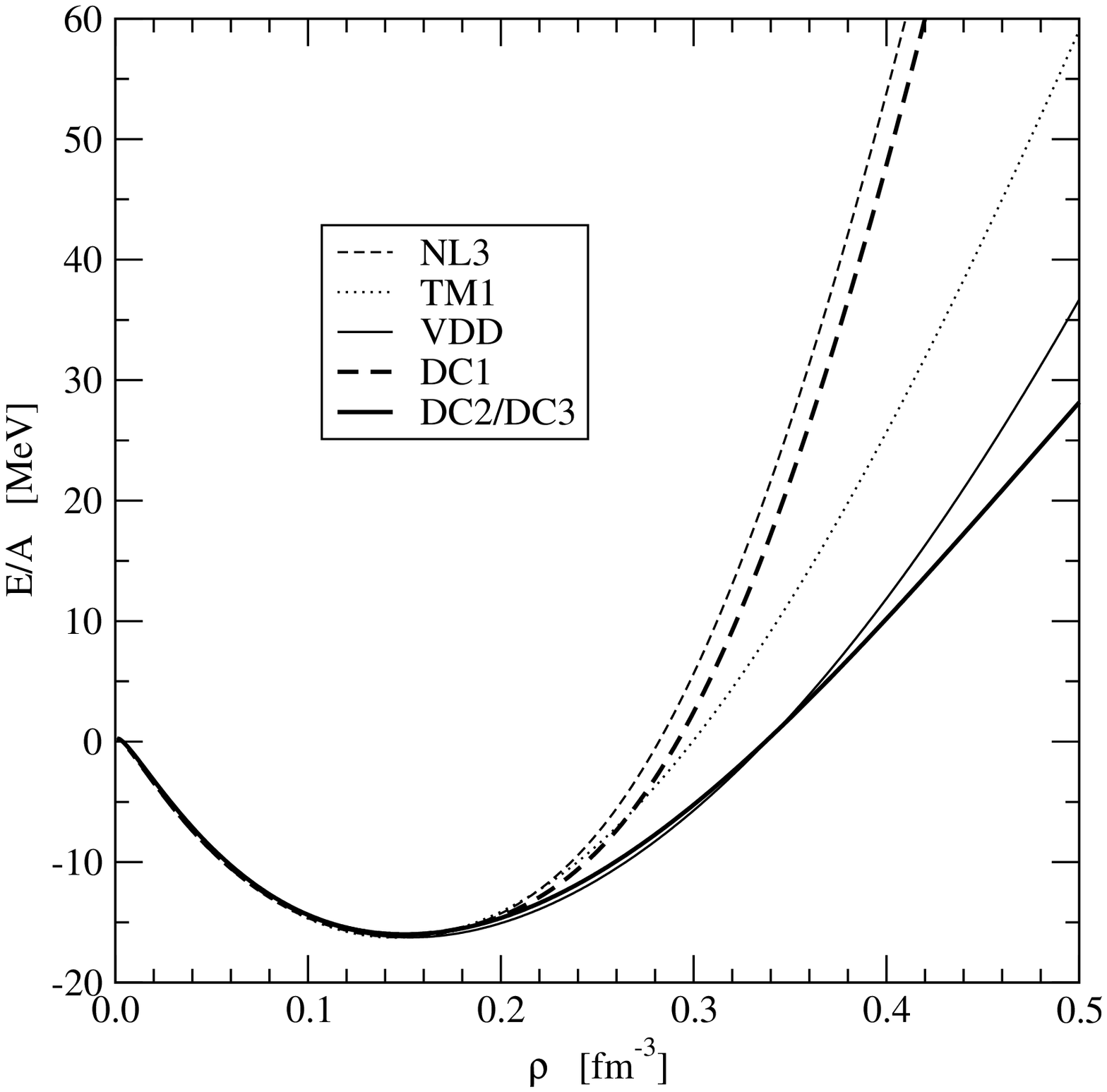}
\caption{\label{fig:eos0} Binding energy per nucleon for symmetric
nuclear matter as a function of the nucleon density in various
relativistic mean field parametrizations.}
\end{figure}

It is instructive to compare various properties of symmetric nuclear
matter and neutron matter in the DC models with other typical relativistic
mean field models, in particular with standard non-linear parametrizations
and models with density-dependent couplings.
Here we chose the parametrization NL3 \cite{Lal97} which contains tertic and
quartic self-interactions of the $\sigma$-meson field and the set
TM1 \cite{Sug94} which takes also self-interactions of the $\omega$-meson into
account. The VDD model \cite{Typ99} 
assumes a dependence of meson-nucleon couplings 
on the vector density that generates so-called ``rearrangement''
contributions in the self-energies leading to a non-trivial
density dependence. These parametrizations
were obtained by fits to properties of finite nuclei and nuclear matter.

\begin{figure}
\includegraphics[width=13cm]{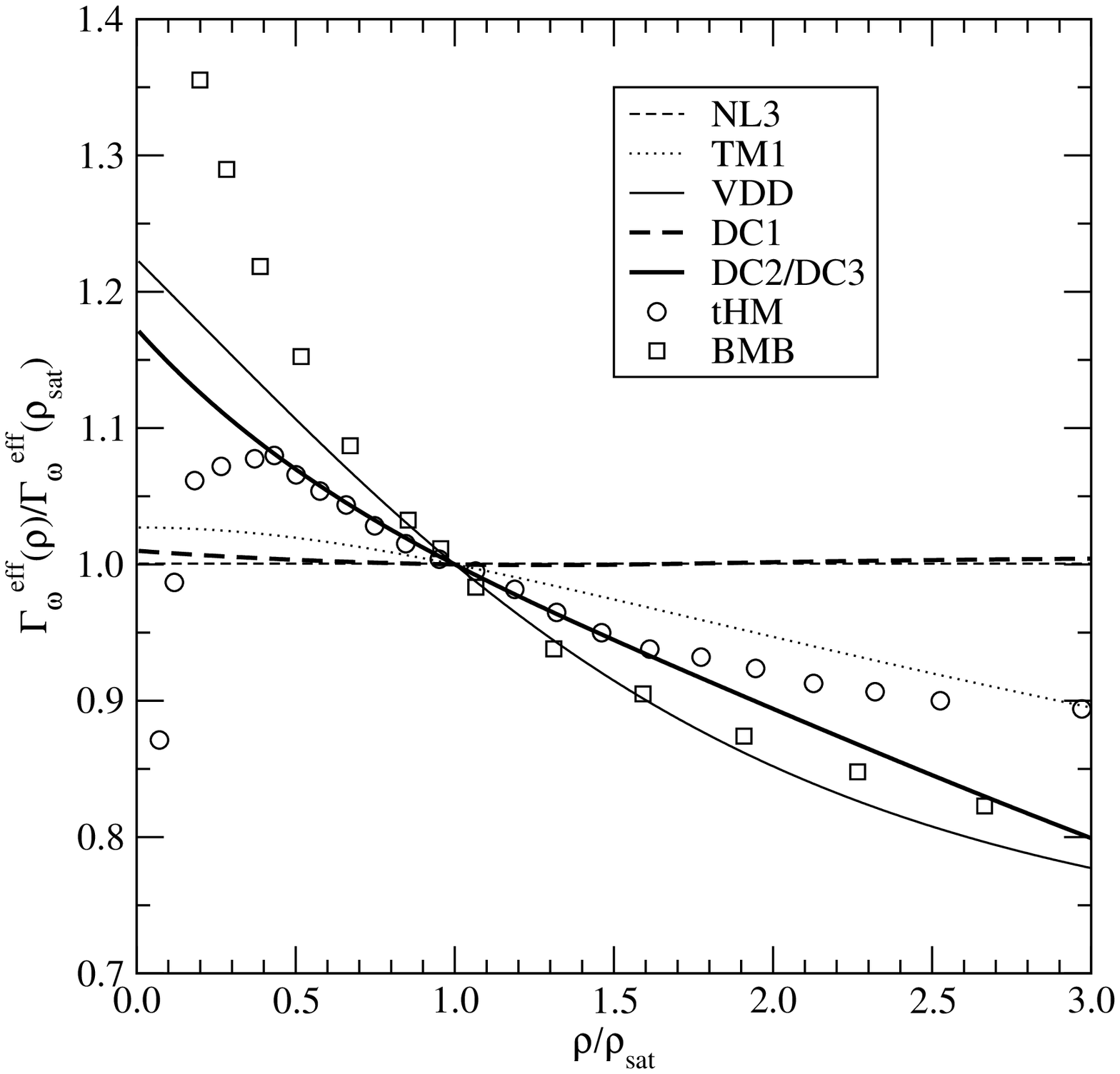}
\caption{\label{fig:gameffv}  Normalized effective coupling constant
of the $\omega$-meson in symmetric nuclear matter 
as a function of the normalized
nucleon density in various relativistic mean field parametrizations and
Dirac-Brueckner calulations.}
\end{figure}

\begin{figure}
\includegraphics[width=13cm]{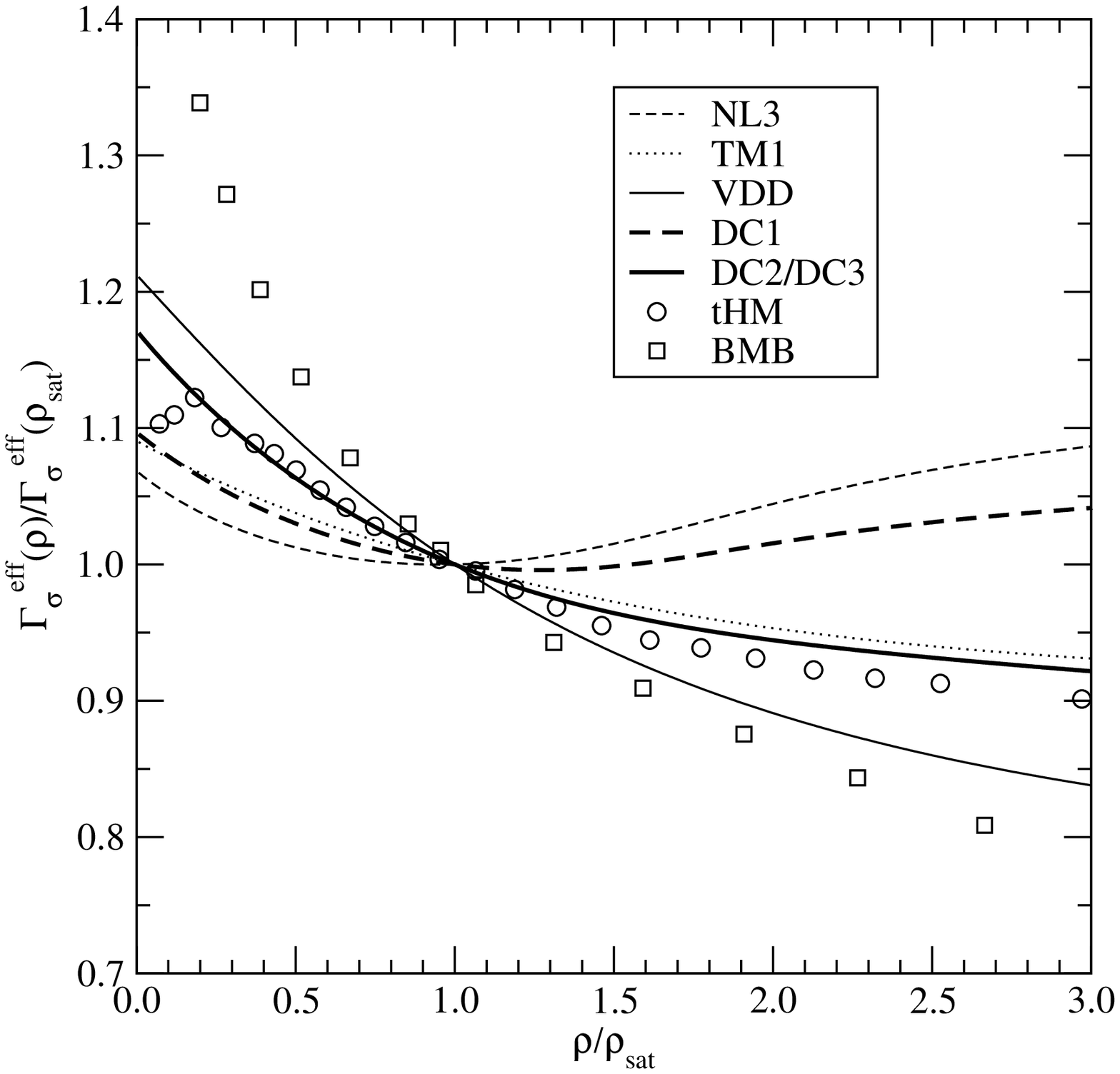}
\caption{\label{fig:gameffs} Normalized effective coupling constant
of the $\sigma$-meson in symmetric nuclear matter 
as a function of the normalized
nucleon density in various relativistic mean field parametrizations and
Dirac-Brueckner calulations.}
\end{figure}

In Figure \ref{fig:eos0} the binding energy of symmetric nuclear
matter is shown for different models as a function of
the density. Below saturation density all parametrizations 
lead to almost identical results, but at higher densities
there are characteristic differences. The DC1 model without 
energy dependent effective self energies is very
similar to the NL3 parametrization with the non-linear
self-interaction of the $\sigma$ meson. The effective density-dependence
of the meson masses in the DC1 model with $\Gamma_{\sigma}^{(2)}
\gg \Gamma_{\omega}^{(2a)}$ apparently leads to an effective density
dependence of the scalar and vector self-energies that is
comparable to the standard non-linear model. 
This is easily observed by comparing the effective coupling
constants in symmetric nuclear matter. They are extracted from
the self-energies according to
\begin{equation}
 \Gamma_{\sigma}^{\rm eff} = m_{\sigma}
 \left( \frac{\Sigma}{\varrho_{s}}\right)^{\frac{1}{2}}
 \qquad \mbox{and} \qquad
 \Gamma_{\omega}^{\rm eff} = m_{\omega}
 \left( \frac{\Sigma_{0}}{\varrho}\right)^{\frac{1}{2}} \: .
\end{equation}
The effective coupling constants as shown in Figures
\ref{fig:gameffv} and \ref{fig:gameffs} are normalized to one
at saturation density of the respective parametrization
and they are independent of the meson masses.
In the NL3 parametrization
the effective coupling of the $\omega$-meson is independent of the density
and it is almost density independent
in the DC1 model. On the other hand, the effective coupling of the
$\sigma$-meson shows in both models
an increase for large densities. This behaviour is in contrast
to all other RMF models.

The TM1 parametrization
achieves a slightly softer equation of state for
symmetric nuclear matter at high densities than the NL3 and DC1 models,
respectively,
and the effective $\omega$ coupling
decreases significantly with increasing density.
In this model a quartic self-coupling of the $\omega$-meson 
are taken explicitly into account in the Lagrangian density. It
gives rise to a $\omega_{0}^{3}$ contribution in the meson field
equation that becomes important at high densities and consequently 
reduces the vector self-energy.

The VDD model shows an even softer
equation of state at high densities as compared to the above
parametrizations. This behaviour is determined by the functional
form of the density dependence of the coupling functions, which was chosen
in order to describe effective coupling constants
extracted from Dirac-Brueckner calculations of nuclear matter. 
In Figures \ref{fig:gameffv} und \ref{fig:gameffs} 
the corresponding results of two DB calculations (BMB with Bonn B potential
\cite{Bro90}, tHM
\cite{tHM87})
are shown for comparison. Here, the strong decrease of the
couplings with the density is evident although there is a considerable
difference between the DB calculations. 
The DC2 and DC3 models,
which are only different with respect to their isovector 
properties, also exhibit a reduction of the effective couplings
similar to the VDD parametrization and the DB results. 
The derivative couplings in the Lagrangian 
generate an energy and density dependence 
of the self energies that leads to
a significant decrease of the effective couplings.
Consequently, the binding energy per nucleon rises much
slower with density than in the NL3, DC1 and TM1 parametrizations.

\begin{figure}
\includegraphics[width=13cm]{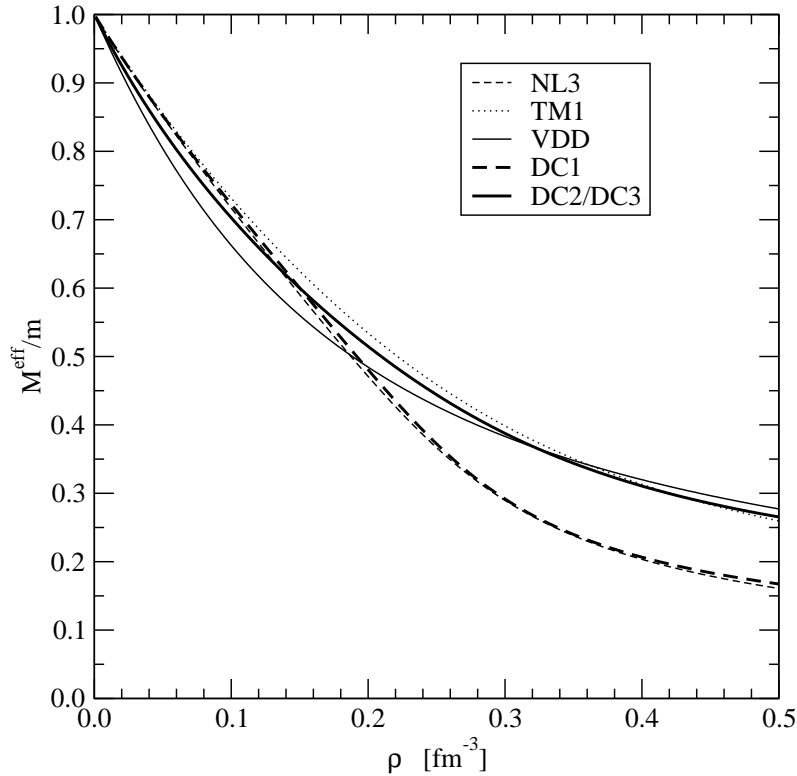}
\caption{\label{fig:meff} Relativistic effective nucleon mass
in symmetric nuclear matter
in units of the nucleon rest mass as a function of the nucleon 
density in various relativistic mean field parametrizations.}
\end{figure}

The density dependence of the effective $\sigma$-meson coupling
and of the scalar self-energy is directly related to 
the decrease of the relativistic effective mass 
of the nucleon $M^{\rm eff}$ which is
shown in Figure \ref{fig:meff} for symmetric nuclear matter. 
We recall that the effective mass of the DC models is fixed to a
value of $0.6$ at $\varrho_{\rm sat}$.
There a two distinctly different
groups, especially at high densities. 
The parametrizations NL3 and DC1 show a very similar behaviour
as for the equation of state. The effective mass of these sets is
substantially smaller than the mass of the other parametrizations 
at high densities and the
curve exhibits less curvature. 
In symmetric nuclear matter the effective
nucleon masses in the DC2 and DC3 model are identical but depend
except on the density of the medium also on the momentum of the nucleon.
In Figure \ref{fig:meff} the effective mass of nucleons
in the DC2/DC3 parametrizations is shown
for the Fermi momentum $p^{F}$. 
The effective mass in the VDD model
is smaller than in the other models at saturation density which is
reflected by the too large spin-orbit splitting in finite nuclei \cite{Typ99}
but the density dependence is more similar to the TM1, DC2 and DC3 cases.

\begin{figure}
\includegraphics[width=13cm]{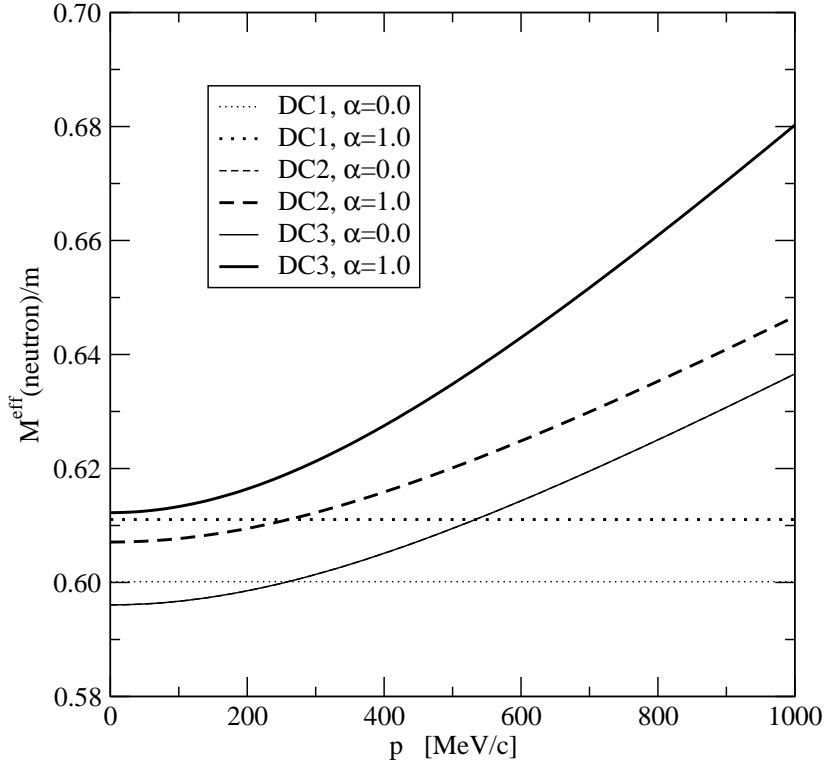}
\caption{\label{fig:meffa} Momentum dependence of the 
neutron effective mass in units of the nucleon mass for the DC
parametrizations in symmetric nuclear matter (thin lines) and
neutron matter (thick lines) at $\varrho=0.150$~fm${}^{-3}$.
(The curves for $\alpha=0.0$ in the DC2 and DC3 model are identical.)}
\end{figure}

In Fig.~\ref{fig:meffa} we show the momentum dependence of the
neutron effective mass of different models for different asymmetries.
The effective nucleon mass $M^{\rm eff}$
in the DC models is fixed to $0.6$~m at 
saturation density $\varrho_{\rm sat}=0.150$~fm${}^{-3}$ for
nucleons with the Fermi momentum in symmetric nuclear matter. 
The scalar self energy
in the DC1 parametrization does not depend on the energy and 
therefore the effective
mass is momentum independent. It increases 
only slightly with the 
neutron-proton asymmetry $\alpha$ at constant density.
In the DC2 and DC3 models $M^{\rm eff}$ increases with
the momentum and with the asymmetry. 
For $\alpha=0$ the effective mass is identical
for DC2 and DC3 but in asymmetric nuclear
matter the absolute value and the momentum dependence of $M^{\rm eff}$
are different.
This effect is caused by the different values of $\Gamma_{\rho}$ and
the non-zero $\Gamma_{\rho}^{(1)}$ in the DC3 case.
The momentum dependence becomes 
stonger at higher densities.

\begin{figure}
\includegraphics[width=13cm]{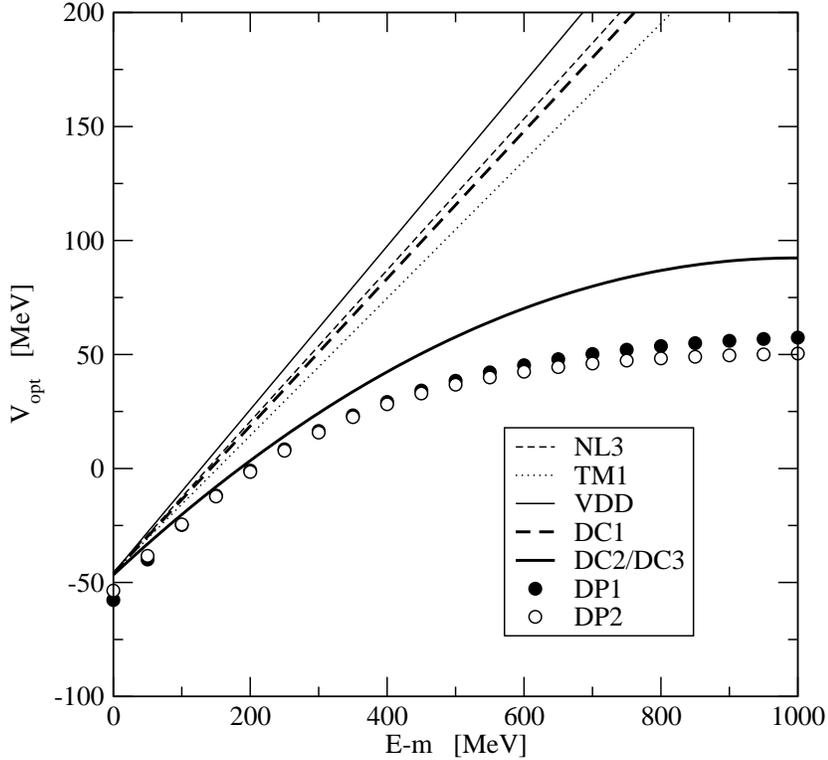}
\caption{\label{fig:vopt10}  Dependence of the 
Schr\"{o}dinger equivalent optical
potential on the nucleon energy 
in symmetry nuclear matter at $\varrho = 0.150$~fm${}^{-3}$ in various
relativistic mean field models and in two fits (DP1, DP2)
of proton-nucleus scattering in Dirac phenomenology.}
\end{figure}

\begin{figure}
\includegraphics[width=13cm]{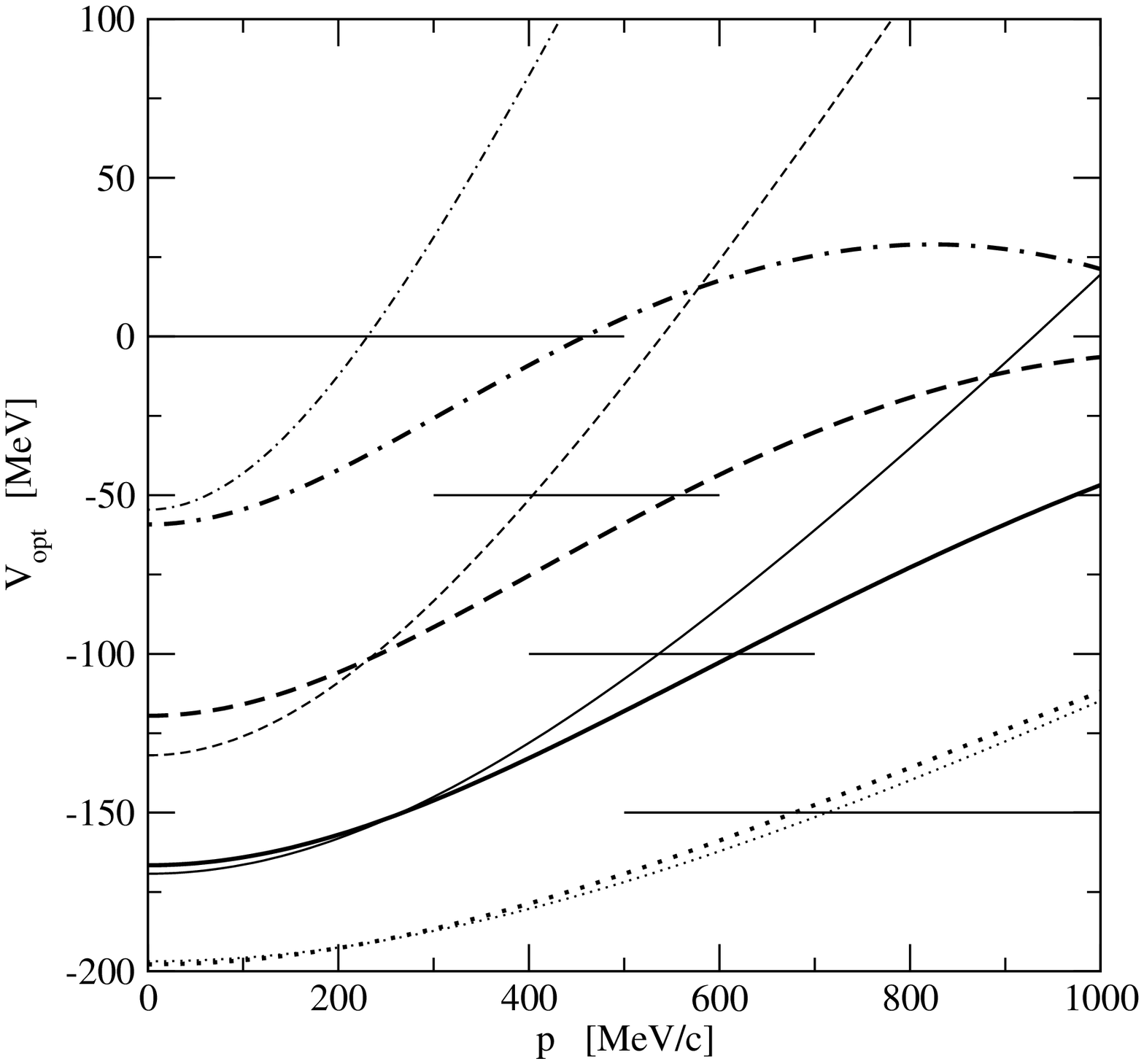}
\caption{\label{fig:voptp}  Dependence of the 
Schr\"{o}dinger equivalent optical
potential on the nucleon momentum 
in symmetric nuclear matter at various densities for the parametrizations
DC1 (thin lines) and DC2/DC3 (thick lines). The potentials at
$\varrho = 0.5 \varrho_{\rm sat}$ (dotted lines), 
$\varrho = 1.0 \varrho_{\rm sat}$ (solid lines), 
$\varrho = 1.5 \varrho_{\rm sat}$ (dashed lines),
and $\varrho = 2.0 \varrho_{\rm sat}$ (dot-dashed lines)
are shifted downwards by 150~MeV, 100~MeV, 50~MeV and 0~MeV,
respectively. Horizontal lines mark the range in the momentum
where the potential crosses zero.}
\end{figure}

The main difference between the DC2/DC3 parametrizations and the other models
becomes apparent when the Schr\"{o}dinger equivalent
optical potential is studied, which is shown in 
Fig.~\ref{fig:vopt10} for $\varrho = \varrho_{\rm sat}$ as
a function of the nucleon energy. In all models
without energy dependent self-energies the optical potential 
by construction rises
linearly with the energy in contrast to
the real part of the optical potential extracted from Dirac phenomenology
\cite{Ham90,Coo93}.
This is clearly seen in Figure \ref{fig:vopt10}.
At energies $E-m$ below approx. 200~MeV the optical potential
is attractive. It higher energies it becomes repulsive.
The linear increase of the RMF models leads to a substantial
overestimate of the optical potential at energies above a few
hundred MeV. Only the DC2 and DC3 models
show a significant reduction of the repulsive potential at high energies,
even though it is still larger than the experimental results
from Dirac phenomenology.
A fit of the parameters in the DC model with
less constraints on the coupling constants than in the DC2/DC3 sets
can certainly improve this
agreement but one has to consider that the phenomenological scalar and
vector potentials have an imaginary part which is absent in the DC model.
Therefore, a comparison 
of the optical potentials at high energies has to viewed
with caution. However, the DC model represents a significant 
qualitative improvement
as compared to standard RMF models.

\begin{figure}
\includegraphics[width=13cm]{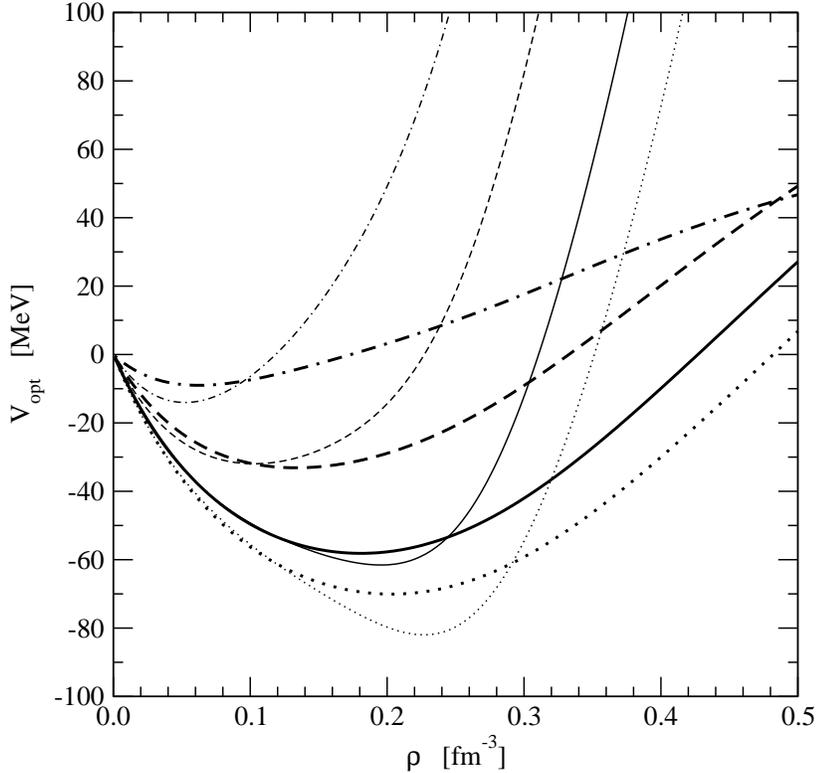}
\caption{\label{fig:voptrh}  Dependence of the Schr\"{o}dinger 
equivalent optical
potential on the density
in symmetric nuclear matter  for the parametrizations
DC1 (thin lines) and DC2/DC3 (thick lines) at nucleon momenta of
$p=0$~MeV (dotted lines), 
$p=200$~MeV (solid lines), 
$p=400$~MeV (dashed lines),
and $p=600$~MeV (dot-dashed lines).}
\end{figure}

\begin{figure}
\includegraphics[width=13cm]{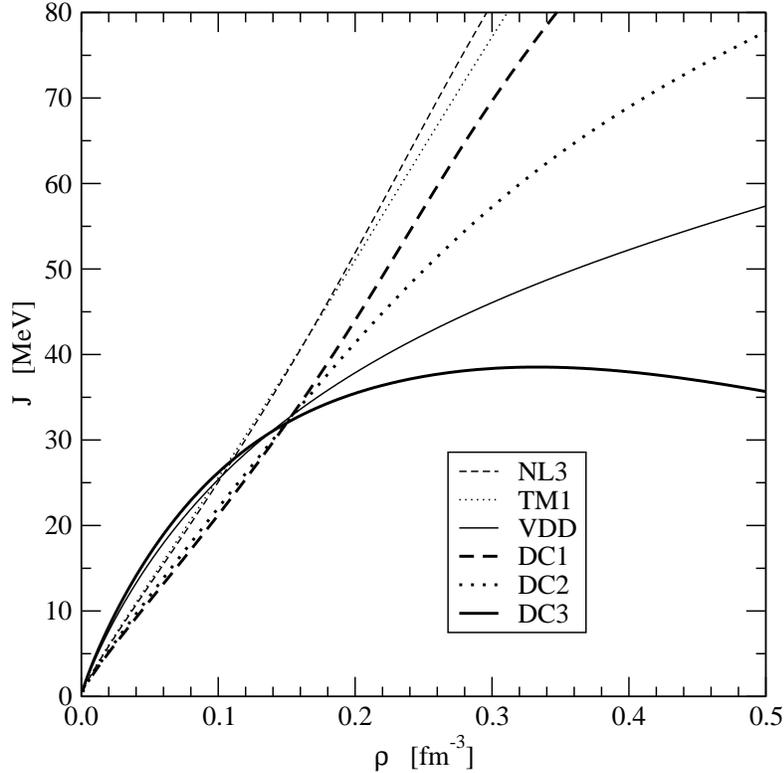}
\caption{\label{fig:sym}  Dependence of the symmetry energy for symmetric
nuclear matter on the nucleon density in various
relativistic mean field parametrizations.}
\end{figure}

The DC models also makes a prediction on the momentum
dependence of the optical potential at different medium densities.
In Figure \ref{fig:voptp} the results of the parametrization 
DC1 are compared to
the DC2/DC3 parametrizations at four nuclear matter densities between
one half and twice the saturation density. The DC1 parameter set
exhibits the typical behaviour of a RMF model with energy-independent
self-energies. The optical potential rises fast with increasing
nucleon momentum and becomes repulsive 
at progressively smaller momenta as the
density of the medium increases. The DC2/DC3 models with 
momentum-dependent self-energies shows a similar trend as the DC1 model
at small densities but at higher densities the optical potential
is much less repulsive than in the DC1 model. The shift in the
zero of $V_{\rm opt}$ is less pronounced and the nontrivial
momentum dependence generates a considerable curvature. 
The density-dependence of the optical potential for constant nucleon 
momentum is shown in Fig.~\ref{fig:voptrh}
and exhibits significant differences when the set DC1 is compared
to the sets DC2/DC3. 
The optical potential vanishes at zero density independent
of the nucleon momentum. It becomes smaller with increasing density 
and reaches a minimum before it rises again. The minimum is deeper
for smaller momenta. At lower densities the parametrizations DC1 and
DC2/DC3 display a more or less similar behaviour but at high densities
the optical potential in the model without momentum dependent self-energies
strongly increases and becomes very repulsive 
whereas the slope of $V_{\rm opt}$ in the other model 
is much smaller and the repulsion sets in much later.


\begin{figure}
\includegraphics[width=13cm]{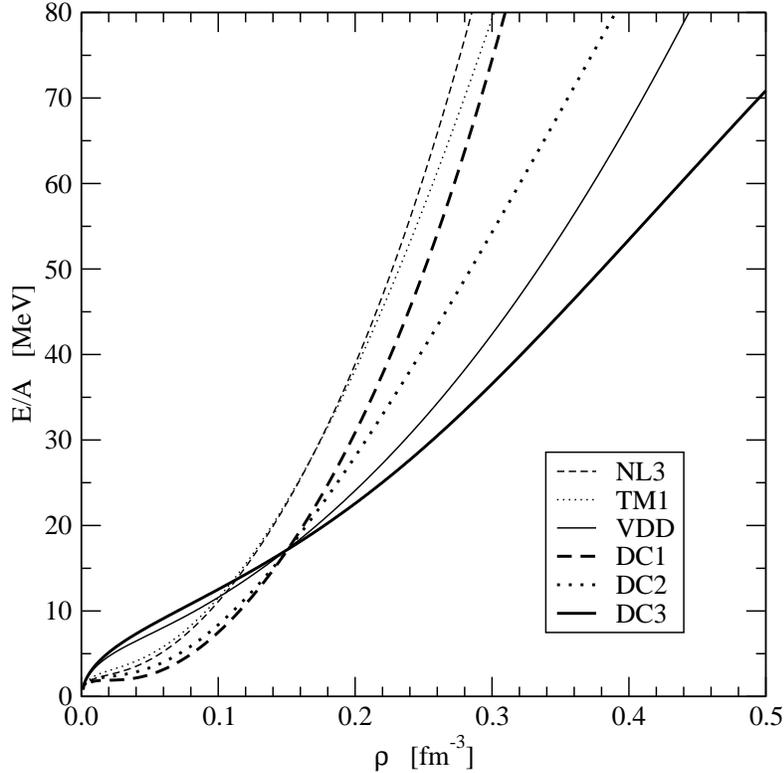}
\caption{\label{fig:eos1} Binding energy per nucleon for 
neutron matter as a function of the neutron density in various
relativistic mean field parametrizations.}
\end{figure}

The various parametrizations also
exhibit a clearly distinct behaviour in the
density dependence of the symmetry energy
which is presented in Figure \ref{fig:sym}. 
The sets NL3, TM1, DC1, DC2
show an almost linear increase of $J$ over a large range of 
densities. In all these models
the coupling of the nucleons to the $\rho$-meson
is described by only one parameter, the constant $\Gamma_{\rho}$ of
the minimal coupling.
In the VDD model, the curve bends significantly with a smaller slope
at saturation density because $\Gamma_{\rho}$ is a decreasing
function with density.
The effect is even larger in
the DC3 parametrization where the symmetry energy derivative $L$ at
saturation density was fitted to a smaller but more realistic
value than in the other relativistic models. 
Comparing only the parametrizations NL3, TM1, and VDD, which were fitted
to properties of finite nuclei, one notices that the symmetry energy
at saturation density
of the NL3 and TM1 models is much larger than in the VDD model
but that $J$ is very similar around a density of $0.11$~fm${}^{-3}$.

The density dependence of the symmetry energy directly correlates 
with the slope of the equation of state for neutron matter which is shown
in Figure \ref{fig:eos1}. The parametrizations NL3, TM1, DC1, and
DC2 show a much stiffer equation of state than the density-dependent
VDD model and the DC3 model with the smaller symmetry energy derivative $L$.
The difference of the models is especially apparent at small densities.
In Refs.\ \cite{Bro00,Typ01} it was
shown that the slope of the neutron matter equation of state
is directly correlated with the neutron skin thickness of finite nuclei, 
i.e.\ the difference between  neutron and proton radii.
This relation holds in both
nonrelativistic Skyrme Hartree-Fock models and relativistic mean-field
models.  Experiments to
determine neutron radii of finite nuclei with higher precision than
available today can lead to better
constraints on the model parameters and to better predictions for 
the neutron matter equation of state.
It can be expected that the DC3 model with the 
smallest slope will also give a small neutron skin thickness similar to
several Skyrme parametrizations when applied to calculations of finite
nuclei.

\section{Summary and outlook}
\label{Sec5}

The DC model is an extension of standard relativistic quantum hadronic models
in order to generate an effective momentum dependence of the self energies
in the mean field approximation. This is achieved by introducing 
couplings of the meson fields to derivatives of the nucleon field in
the Lagrangian density. The new contributions lead to additional source terms
in the meson field equations and to a density dependence
of the effective meson masses. The effective interaction in the DC model 
is both medium and state dependent.
The self energies in the Dirac equation are differential operators
with a nontrivial density dependence. The effective mass depends
in general on the density, the momentum and the neutron-proton asymmetry.

We applied the DC model to nuclear matter and
developed three parametrizations for
the coupling constants under different conditions.
The set DC1 without momentum dependent self energies
is similar to previous nonlinear RMF models.
It displays rather stiff equations of state for symmetric
nuclear matter and neutron matter. 
The sets DC2 and DC3 lead to energy dependent self energies
The equation of state becomes softer and is comparable to the RMF model
with density dependent couplings. The 
Schr\"{o}dinger equivalent optical potential is much less
repulsive than in standard RMF parametrizations at high nucleon energies.
It shows an energy dependence more like optical potentials extracted
from Dirac phenomenology. The DC model is also flexible to adjust
isovector properties which affects the neutron matter equation of state.

If the nucleon momenta are too large the assumption of
a linear energy dependence of the self energies is no longer valid.
In this case the DC model can no longer be applied.
This deficiency could be improved by introducing couplings of the
meson fields to higher derivatives of the nucleon fields.
Since the medium density determines the Fermi momentum
there is also a limitation of the model to not too large densities
but the Fermi momentum increases only slowly with the density.

Because of the limited number of constraints from
nuclear matter properties
the parameter space of the DC model was not fully explored.
It is worthwhile to apply the model to finite nuclei and to
study the effects of the new coupling vertices in this case.
Also important is the comparison to extensive experimental data
on nucleon scattering.
The parametrizations developed here can be used as a
reasonable starting point for a further refinement.

The approach of the DC model with fixed coupling constants but
generalized interaction vertices in the Lagrangian density is
in the spirit of non-linear RMF models. Alternatively, one can
think of an extension
of models with density dependent meson-nucleon vertices
by introducing a dependence of the coupling functions 
on derivative densities. We currently also investigate the possibilities
of such an approach and results will be reported in the future.

\acknowledgments

Support for this work was provided in part by grant LMWolT from
GSI and in part from U.S. National Science
Foundation grant No. PHY-0070911.

\appendix

\section{Explicit expressions of densities}
\label{appa}

The conserved current densities 
in the field equations of the $\omega$- and $\rho$-meson
are given as
\begin{eqnarray}
 J_{\mu} 
 & = & [( 1 + \mathcal{W} )g_{\mu \nu} + \mathcal{X}_{\mu \nu}] 
   j^{\nu}  - \mathcal{T}_{\mu}  \varrho_{s}
 + \vec{\mathcal{W}} \cdot \vec{\jmath}_{\mu} 
 - \vec{\mathcal{T}}_{\mu} \cdot \vec{\varrho}_{s}
 \\ 
 \vec{J}_{\mu} 
 & = & [( 1 + \mathcal{W} )g_{\mu \nu} + \mathcal{X}_{\mu \nu}] 
   \vec{\jmath}^{\nu}  - \mathcal{T}_{\mu}  \vec{\varrho}_{s}
  \\ \nonumber & & 
   + \vec{\mathcal{W}} j_{\mu} 
   - i \vec{\mathcal{W}} \times \vec{\jmath}_{\mu}
   - \vec{\mathcal{T}}_{\mu} \varrho_{s} 
   + i \vec{\mathcal{T}}_{\mu} \times \vec{\varrho}_{s} 
\end{eqnarray}
and
depend on the usual scalar densities and vector densities.
The derivative scalar densities 
\begin{eqnarray} 
 \varrho_{s}^{D} & = & 
 \varrho_{s}^{d}  - \Gamma_{\omega} \omega_{\mu} j^{\mu}
 - \Gamma_{\rho} \vec{\rho}_{\mu} \cdot \vec{\jmath}^{\mu}
 = g^{\mu \nu} t^{D}_{\mu \nu}
 \\  
  \vec{\varrho}_{s}^{D} & = & 
 \vec{\varrho}_{s}^{d}  - \Gamma_{\omega} \omega_{\mu} \vec{\jmath}^{\mu}
 - \Gamma_{\rho} \vec{\rho}_{\mu}  j^{\mu}
 + i \Gamma_{\rho} \vec{\rho}_{\mu} \times \vec{\jmath}^{\mu}
 = g^{\mu \nu} \vec{t}^{D}_{\mu \nu}
\end{eqnarray}
with 
\begin{eqnarray} 
 \varrho_{s}^{d} & = &
 \frac{1}{2} \langle \left[ \bar{\psi} \gamma^{\mu} i \partial_{\mu} \psi 
 - ( i \partial_{\mu} \bar{\psi}) \gamma^{\mu} \psi \right] \rangle
 = g^{\mu \nu} t^{d}_{\mu \nu}
 \\ 
  \vec{\varrho}_{s}^{d} & = &
 \frac{1}{2} \langle \left[ \bar{\psi} \gamma^{\mu} 
 i \partial_{\mu} \vec{\tau} \psi 
 - ( i \partial_{\mu} \bar{\psi}) \gamma^{\mu} \vec{\tau} \psi \right] \rangle
 = g^{\mu \nu} \vec{t}^{d}_{\mu \nu}
\end{eqnarray}
are obtained from a contraction of the
derivative tensor densities
\begin{eqnarray}
 t^{D}_{\mu \nu} & = &
 \frac{1}{2} \langle \left[ \bar{\psi} \gamma_{\mu} i D_{\nu} \psi 
 + \overline{(iD_{\nu}\psi)} \gamma_{\mu} \psi \right] \rangle
 = t^{d}_{\mu \nu}  - \Gamma_{\omega} \omega_{\nu} j_{\mu}
 - \Gamma_{\rho} \vec{\rho}_{\nu} \cdot \vec{\jmath}_{\mu}
 \\
  \vec{t}^{D}_{\mu \nu} & = &
 \frac{1}{2} \langle \left[ \bar{\psi} \gamma_{\mu} i D_{\nu} \vec{\tau} \psi 
 + \overline{(iD_{\nu}\psi)} \gamma_{\mu} \vec{\tau} \psi \right] \rangle
 = \vec{t}^{d}_{\mu \nu}  - \Gamma_{\omega} \omega_{\nu} \vec{\jmath}_{\mu}
 - \Gamma_{\rho} \vec{\rho}_{\nu} j_{\mu}
 + i \Gamma_{\rho} \vec{\rho}_{\nu} \times \vec{\jmath}_{\mu}
\end{eqnarray}
where
\begin{eqnarray}
 t^{d}_{\mu \nu} & = &
 \frac{1}{2} \langle \left[ \bar{\psi} \gamma_{\mu} i \partial_{\nu} \psi 
 -(i\partial_{\nu}\bar{\psi}) \gamma_{\mu} \psi \right] \rangle
 \\
 \vec{t}^{d}_{\mu \nu} & = &
 \frac{1}{2} \langle \left[ \bar{\psi} \gamma_{\mu} 
 i \partial_{\nu} \vec{\tau} \psi 
 -(i\partial_{\nu}\bar{\psi}) \gamma_{\mu} \vec{\tau} \psi \right] \rangle
\end{eqnarray}
with the metric tensor $g^{\mu \nu}$.
For the derivative current densities the expressions
\begin{eqnarray}
 j^{D}_{\mu} & = &
 j^{d}_{\mu}  - \Gamma_{\omega} \omega_{\mu} \varrho_{s}
 - \Gamma_{\rho} \vec{\rho}_{\mu} \cdot \vec{\varrho}_{s}
 \\ 
 \vec{\jmath}^{D}_{\mu} & = &
 \vec{j}^{d}_{\mu}  - \Gamma_{\omega} \omega_{\mu} \vec{\varrho}_{s}
 - \Gamma_{\rho} \vec{\rho}_{\mu} \varrho_{s}
 + i \Gamma_{\rho} \vec{\rho}_{\mu} \times \vec{\varrho}_{s}
\end{eqnarray}
with
\begin{eqnarray}
 j^{d}_{\mu} & =  &
 \frac{1}{2} \langle \left[ \bar{\psi}  i \partial_{\mu} \psi 
 - ( i \partial_{\mu} \bar{\psi})  \psi \right] \rangle 
 \\
 \vec{j}^{d}_{\mu} & =  &
 \frac{1}{2} \langle \left[ \bar{\psi}  i \partial_{\mu} \vec{\tau} \psi 
 - ( i \partial_{\mu} \bar{\psi}) \vec{\tau} \psi \right] \rangle \: .
\end{eqnarray}
are found.

\section{Normalization of spinors}
\label{appb}

The spinors (\ref{spinsol}) in the solutions of the Dirac equation are 
normalized according to
\begin{equation}
  \bar{u}( \vec{p}, \sigma, \tau ) \Gamma_{0\tau} 
  u( \vec{p}, \sigma^{\prime}, \tau^{\prime} ) 
  =   \delta_{\sigma \sigma^{\prime}} \delta_{\tau \tau^{\prime}} 
\end{equation}
which guarantees that the zero component of the current density $J_{\mu}$ 
can be interpreted as the  baryon density. 
Correspondingly, the relations
\begin{eqnarray} 
 \bar{u}( \vec{p}, \sigma, \tau ) \gamma_{0} 
  u( \vec{p}, \sigma^{\prime}, \tau^{\prime} ) 
 & = &  \frac{\tilde{E}_{\tau}\delta_{\sigma \sigma^{\prime}} 
  \delta_{\tau \tau^{\prime}}}{(1+W_{\tau}+X_{00 \tau})\tilde{E}_{\tau}
 -T_{0 \tau}\tilde{M}_{\tau}}
 \\ \nonumber 
 & = &  \frac{\delta_{\sigma \sigma^{\prime}} 
  \delta_{\tau \tau^{\prime}}}{(1+W_{\tau}+X_{00 \tau})(1-B_{\tau}^{2})}
 \left( 1 + \frac{A_{\tau}B_{\tau}}{\sqrt{A_{\tau}^{2}+\tilde{p}_{\tau}^{2}}} 
 \right)
\end{eqnarray}
and
\begin{eqnarray} 
 \bar{u}( \vec{p}, \sigma, \tau ) 
 u( \vec{p}, \sigma^{\prime}, \tau^{\prime} ) & = &
 \frac{\tilde{M}_{\tau}\delta_{\sigma \sigma^{\prime}} 
  \delta_{\tau \tau^{\prime}}}{(1+W_{\tau}+X_{00 \tau})\tilde{E}_{\tau}
 -T_{0 \tau}\tilde{M}_{\tau}}
 \\ \nonumber & = & 
 \frac{\delta_{\sigma \sigma^{\prime}} 
  \delta_{\tau \tau^{\prime}}}{(1+W_{\tau}+X_{00 \tau})(1-B_{\tau}^{2})}
 \left( B_{\tau} + \frac{A_{\tau}}{\sqrt{A_{\tau}^{2}+\tilde{p}_{\tau}^{2}}} 
 \right)
\end{eqnarray} 
for the usual scalar and vector densities are found. 
In the calculation of the scalar derivative density also
the relation
\begin{equation} \label{ugpu}
 \bar{u}( \vec{p}, \sigma, \tau ) 
 \vec{\gamma} \cdot \vec{\tilde{p}}_{\tau} 
 u( \vec{p}, \sigma^{\prime}, \tau^{\prime} )  = 
 \frac{\vec{\tilde{p}}_{\tau}^{2}\delta_{\sigma \sigma^{\prime}} 
  \delta_{\tau \tau^{\prime}}}{(1+W_{\tau}+X_{00 \tau})\tilde{E}_{\tau}
 -T_{0 \tau}\tilde{M}_{\tau}} 
\end{equation}
will be needed.


\end{document}